\documentclass[12pt]{article}
\def\baselinestretch{0.98}
\usepackage{cite}
\usepackage{multirow}
\usepackage{amssymb, amsmath, amsopn, amsthm}
\usepackage{bm,amsfonts,array,calc,rotating}
\usepackage{epsfig}
\usepackage{graphicx}
\usepackage{color}
\usepackage{epic}

\definecolor{orange}{cmyk}{0,0.5,1,0}

\newcommand{\ra}{\rightarrow}
\newcommand{\va}{\varphi}
\usepackage[T1]{fontenc}

\numberwithin{equation}{section}

\long\def\@makecaption#1#2{%
  \vskip\abovecaptionskip
  \sbox\@tempboxa{{\bf #1:} #2}%
  \ifdim \wd\@tempboxa >\hsize
    {\small\bf #1:} {\small #2}\par
  \else
    \global \@minipagefalse
    \hb@xt@\hsize{\hfil\box\@tempboxa\hfil}%
  \fi
  \vskip\belowcaptionskip}

\font\cmss=cmss10 \font\cmsss=cmss10 at 7pt
\def\IZ{\relax\ifmmode\mathchoice
{\hbox{\cmss Z\kern-.4em Z}}{\hbox{\cmss Z\kern-.4em Z}}
{\lower.9pt\hbox{\cmsss Z\kern-.4em Z}} {\lower1.2pt\hbox{\cmsss
Z\kern-.4em Z}}\else{\cmss Z\kern-.4em Z}\fi}

\def\sqr#1#2{{\vcenter{\vbox{\hrule height.#2pt
 \hbox{\vrule width.#2pt height#1pt \kern#1pt
 \vrule width.#2pt}\hrule height.#2pt}}}}

\begin{document}

\begin{flushright}
\baselineskip=12pt \normalsize
{MIFPA-10-38}\\
\smallskip
\end{flushright}

\begin{center}
\Large {\textbf{On Global Flipped $SU(5)$ GUTs in F-theory}} \\[2cm]
\normalsize  Yu-Chieh Chung
\\[.25in]

\textit{Department of Physics $\&$ Astronomy, Texas A$\&$M University\\ College Station, TX 77843, USA} \\[0.5cm]
\tt \footnotesize ycchung@physics.tamu.edu\\[2.5cm]
\end{center}

\renewcommand{\baselinestretch}{1.5}
\setlength{\baselineskip}{18pt}

\begin{abstract}
We construct an $SU(4)$ spectral divisor and its factorization of
types $(3,1)$ and $(2,2)$ based on the construction proposed in
\cite{Marsano:2010ix}. We calculate the chiral spectra of flipped
$SU(5)$ GUTs by using the spectral divisor construction. The
results agree with those from the analysis of semi-local spectral
covers. Our computations provide evidence for the validity of
the spectral divisor construction and suggest that the standard
heterotic chirality formulae are applicable to the case of
F-theory on an elliptically fibered Calabi-Yau fourfold with no
heterotic dual.
\end{abstract}
\vspace{4cm}


\newpage
\setcounter{page}{1}

\setcounter{footnote}{0}

\pagenumbering{arabic}

\pagestyle{plain}

\section{Introduction}

F-theory \cite{Vafa:1996zn,Vafa:1996yn,Vafa:1996xn} is a
twelve-dimensional geometric version of string theory. The
construction of F-theory is motivated by the $SL(2,\mathbb{Z})$
symmetry in type IIB string theory. The $SL(2,\mathbb{Z})$
symmetry becomes the geometrical reparametrization symmetry of the
torus when the axio-dilaton in type IIB string theory is
identified with the complex modulus of a torus. The
ten-dimensional background of type IIB string theory is lifted to
a twelve-dimensional manifold which admits an elliptic fibration.
Due to the monodromy of $SL(2,\mathbb{Z})$, F-theory can be
regarded as a non-perturbative completion of type IIB string
theory\footnote{See\cite{Denef:2008wq} for a review.}. In
F-theory, it was shown\cite{Bershadsky:1996nh} that the
singularities of elliptic fibers correspond to the gauge groups on
the seven-branes. More precisely, the $A_n$, $D_n$, and $E_n$
singularities of elliptic fibration correspond to $SU(n+1)$,
$SO(2n)$, and $E_n$ gauge groups, respectively. Since F-theory
incorporates the exceptional groups, it is believed to be a
natural framework for model building. Recently, supersymmetric
Grand Unified Theory (GUT) models have been studied extensively in
F-theory framework, in particular, the local version of GUT models
have been explored in \cite{King:2010mq, Chung:2009ib,
Vafa:NoncommutativeandYukawa, Conlon:local02, Conlon:local01,
Randall:localFlavor, Heckman:local02a, Li:local03,
Ibanez:localFlavor02, Ibanez:locaFlavorl01, BHV:2008I, BHV:2008II,
Donagi:2008sl,Donagi:2008ll, Heckman:local01, Heckman:local02,
Heckman:local03, Heckman:localFlavor01, Heckman:localFlavor02,
Heckman:localFlavor03, Nanopoulos:local01, Nanopoulos:local02,
Blumenhagen:local, Chen:2009me, Bourjaily:local01,
Bourjaily:local02}\footnote{See\cite{Heckman:2010bq} for a
review.}. The semi-local and global $SU(5)$ GUTs in F-theory have
been discussed in \cite{Hayashi:2010zp, Cvetic:2010rq,
Blumenhagen:2010ja, Grimm:2009yu, Curio:global,
Collinucci:global03, Collinucci:global02, Watari:global,
Donagi:global, Caltech:global01, Caltech:global02,
Caltech:global03, Blumenhagen:global01, Blumenhagen:global02,
Tartar:globalFlavor01, Tartar:globalFlavor02, Other:global,
Choi:2010nf, Choi:2010su, Dudas:2009hu, Dudas:2010zb,
Grimm:2010ez,Heckman:2010fh}. For the cases of higher rank GUT
groups, global $SO(10)$ GUTs have been studied
in\cite{Chen:2010ts} and semi-local flipped $SU(5)$
GUTs\cite{dimitri, smbarr, AEHN-0} have been constructed
in\cite{Chen:2010tp}. In this paper we mainly focus on flipped
$SU(5)$ GUTs. The purpose of this paper is to promote the
semi-local flipped $SU(5)$ models studied in\cite{Chen:2010tp} to
the global version by using the spectral divisor construction
proposed in\cite{Marsano:2010ix}.

In F-theory, semi-local GUT models can be constructed by using
spectral cover construction\cite{Donagi:2008ll,Donagi:global}. In
particular, one can use $SU(4)$ spectral covers to build flipped
$SU(5)$ models\cite{Chen:2010tp}. We start with an elliptically
fibered Calabi-Yau fourfold $Z_4$ with a base $B_3$ which contains
a divisor $B_2$ where $Z_4$ exhibits an $E_8$ singularity. To
avoid full F-theory on a complicated elliptically fibered
Calabi-Yau fourfold, we adopt a bottom-up approach to construct
models in the decoupling limit, which lead us to consider a
contractible complex surface $B_2$ inside $B_3$ such that we can
reduce full F-theory on $X_4$ to an effective eight-dimensional
supersymmetric gauge theory on $\mathbb{R}^{3,1}\times
B_2$\cite{BHV:2008I,BHV:2008II,Donagi:2008ll,Donagi:2008sl}. To
achieve the decoupling limit, the surface $B_2$ has to be a del
Pezzo surface \cite{delPezzo:01, delPezzo:02}. To obtain the gauge
group $SU(5)\times U(1)_X$, we unfold the $E_8$ singularity
into a $D_5$ singularity corresponding to unbroken $SO(10)$. This
unfolding can be encoded in an $SU(4)$ spectral cover. It was
shown in \cite{Donagi:global, Donagi:2008ll} that the spectral
cover construction naturally encodes the information of the
unfolding $E_8$ singularity and the gauge fluxes. By unfolding an
$E_8$ singularity, we can engineer the singularity of types $D_5$,
$D_6$, $E_6$, and $E_7$ in the Calabi-Yau fourfold $Z_4$. These
operations correspond to the manipulation of the roots of a
$SU(4)$ spectral cover. Generally we need to turn on certain
fluxes to obtain the chiral spectrum. In F-theory, a natural
candidate is the four-form $G$-flux which consists of three-form
fluxes and gauge fluxes. In type IIB theory, these three-form
fluxes produce back-reaction in the background geometry. It was shown in
\cite{Vafa:NoncommutativeandYukawa,Marchesano:2009rz} that the
three-form fluxes induce noncommutative geometric structures and
also modify the texture of the Yukawa couplings. An example of noncommutative geometry is a
fuzzy space, which has been studied
in the context of F-theory in\cite{Heckman:2010pv}. In this
article we shall turn off these three-form fluxes and focus only
on the gauge fluxes. The chirality of the matter fields in the
representations of $SO(10)$ is determined by the traceless cover
fluxes which are $(1,1)$-forms on the spectral covers. To obtain
the gauge group $SU(5)\times U(1)_X$, we turn on a line bundle
associated with $U(1)_X$ to break $SO(10)$ down to $SU(5)\times
U(1)_X$. The spectrum is then determined by the cover fluxes and
$U(1)_X$ fluxes. In this paper we shall focus on the $SU(4)$
spectral cover and also consider the
factorizations of the spectral cover to construct realistic
flipped $SU(5)$ models. For $U(1)_X$ fluxes breaking $SO(10)$ down
to $SU(5)\times U(1)_X$ and numerical models, we refer readers to
\cite{Chen:2010tp} for the details. A brief review of the
semi-local $SU(4)$ spectral cover can be found in section 3.1. The
analysis of the chiral spectrum under $(3,1)$ and $(2,2)$
factorizations can be found in section 4.1.

The spectral cover construction discussed above is semi-local. To
obtain global flipped $SU(5)$ GUT models, we shall use the
spectral divisor construction which has recently been proposed
in\cite{Marsano:2010ix}. This construction is motivated by
heterotic/F-theory duality \footnote{For another construction from
mirror symmetry and the discussion in a global $U(1)$ gauge
symmetry arising from global restrictions of the Tate model, see \cite{Grimm:2010ez} and
references therein.}. In the heterotic string framework, one can
calculate the chirality of matter fields by specifying a line
bundle or its twist $\gamma^{(4)}_H$ on an $SU(4)$ spectral cover.
It turns out the net chirality $N_{\mathbf{r}}$ of matter field in
the representation $\mathbf{r}$ is given
by\cite{Marsano:2010ix,Other:global}
\begin{equation}
N_{\mathbf{r}}=\int_{\Sigma_{\mathbf{r},H}}\gamma^{(4)}_H,
\end{equation}
where $\Sigma_{\mathbf{r},H}$ is the matter curve of
representation $\mathbf{r}$. It was shown in\cite{Friedman:1997yq}
that the data of the spectral cover can be encoded in a $dP_8$
surface. On the other hand, when the Calabi-Yau fourfold admits a
global $K_3$-fibration over $B_2$, the $K3$ fiber degenerates into
two $dP_9$ surfaces glued together along an elliptic curve in the
stable degenerate limit. The elliptic fibration over $B_2$ becomes
the background Calabi-Yau threefold in the dual heterotic string
compactification. Moreover, the spectral cover data of $E_8$
bundles can be encoded in the pair of $dP_9$ surfaces in F-theory
geometry. The subbundles of $E_8$ correspond to some singularities
in $dP_9$ surfaces. In particular, an unbroken $SO(10)$ gauge
group corresponds to a $D_5$ singularity. It turns out that
following heterotic/F-theory duality we can define the dual
spectral divisor in F-theory framework, which encodes the data of the
spectral cover in heterotic theory\cite{Marsano:2010ix}. In
F-theory, the net chirality formula was proposed to be
\begin{equation}
N_{\mathbf{r}}=\int_{\widehat{\Sigma}_{\mathbf{r}}\cdot
p^{\ast}_{\mathcal{D}^{(4)}_F}B_2}{\gamma}^{(4)}_F,\label{general
chirality}
\end{equation}
where ${\gamma}^{(4)}_F$ is the traceless flux on the $SU(4)$
spectral divisor $\mathcal{D}_F^{(4)}$ and
$p^{\ast}_{\mathcal{D}_F^{(4)}}$ is the projective map
$p^{\ast}_{\mathcal{D}_F^{(4)}}: \mathcal{D}_F^{(4)}\ra B_3$. It
was argued\cite{Marsano:2010ix} that this formula is intrinsic in
the sense that it can be applied to the cases of F-theory
compactifications without heterotic duals and that spectral divisor
construction can be regarded as a global completion of the
semi-local spectral cover construction. The case of an $SU(5)$
spectral divisor has been analyzed in\cite{Marsano:2010ix}. In
this article we shall verify this proposal by comparing the
computations of chirality from an semi-local $SU(4)$ spectral
cover with that from an $SU(4)$ spectral divisor. It turns out
that they agree with each other. Our computation provides evidence
to support the validity of the spectral divisor construction. The
detailed construction of the $SU(4)$ spectral divisor can be found
in section 3.2. We also calculate the chirality under $(3,1)$ and
$(2,2)$ factorizations by using the spectral divisor construction.
The results can be found in section 4.2.

The organization of the rest of the paper is as follows: in
section 2, we first briefly review the $SU(4)$ spectral cover
construction and computation of the chiral spectrum in heterotic
string compactification on an elliptically fibered Calabi-Yau
threefold. We then turn to the del Pezzo surface construction for
$SU(4)$ bundles and stable degenerate limits, which are two
important ingredients for heterotic/F-theory duality. We construct
an $SU(4)$ spectral divisor in F-theory motivated by
heterotic/F-theory duality and calculate the chiral spectrum in
the end of section 2. In section 3, we consider the cases of
F-theory compactifications without heterotic duals. We first
briefly review the semi-local $SU(4)$ spectral cover construction
and then turn to constructing an $SU(4)$ spectral divisor. In
section 4, we study $(3,1)$ and $(2,2)$ factorizations of the
$SU(4)$ spectral cover and $SU(4)$ spectral divisor. We also
calculate the chirality induced by traceless fluxes and found
agreement between these two constructions. We summarize and
conclude in section $5$.

\section{Preliminaries}

In this section we shall briefly review the spectral cover
construction in heterotic string. In particular, we shall focus on
the case of an $SU(4)$ spectral cover. We then give an
introduction to heterotic/F-theory duality. In the end of this
section, we construct the dual F-theory spectral
divisor\cite{Marsano:2010ix} motivated by heterotic/F-theory
duality.

\subsection{$SU(4)$ Cover in Heterotic String}

The $\mathcal{N}=1$ four-dimensional effective theory of heterotic
string compactifications\footnote{Here we focus on the case of
$E_8\times E_8$ heterotic string compactificatuion with vanishing
background three-form flux $H$ and with constant dilaton $\phi$.}
is governed by the data $(Z_3,V_1,V_2)$, where $V_1$ and $V_2$ are
vector bundles over a six-dimensional manifold $Z_3$. For
simplicity, we only focus on one of the vector bundles, denoted by
$V$ whose structure group is $G$. Supersymmetry requires that
$Z_3$ be a Calabi-Yau threefold and that $V$ admit a connection
satisfying the Hermitian Yang-Mills equations\cite{GSW}
\begin{equation}
F_{ab}=F_{\bar a\bar b}=0,\;\;\;\;\;g^{a\bar b}F_{a\bar
b}=0,\label{HYM}
\end{equation}
where $g$ and $F$ are a metric of $Z_3$ and curvature of the
connection, respectively. The unbroken gauge group of the
four-dimensional effective theory is then the commutant of $G$ in
$E_8$. To obtain an unbroken $SO(10)$ gauge group, we shall focus
on the case of $G=SU(4)$. It is an extremely difficult task to
construct solutions of the Hermitian Yang-Mills equations
Eq.~(\ref{HYM}) for manifolds of dimension greater than one.
However, it was proven in \cite{Donalson,UY01,UY02} that there is
a one-to-one correspondence between the solutions of the Hermitian
Yang-Mills equations and the construction of stable holomorphic
vector bundles over the same complex manifold\footnote{Let $E$ be
a holomorphic vector bundle over $Z_3$ and $J_{Z_3}$ be a K\"ahler
form of $Z_3$. The slope $\mu (E)$ is defined by $\mu
(E)=\frac{\int_{Z_3}c_{1}(E)\wedge J_{Z_3}\wedge J_{Z_3}}{{\rm
rk}(E)}$. The vector bundle $E$ is (semi)stable if for every
subbundle or subsheaf $\mathcal{E}$ with ${\rm
rk}(\mathcal{E})<{\rm rk}(E)$, the inequality $ \mu
(\mathcal{E})<(\leqslant)\mu (E){\label{Stability}}$ holds. Assume
that $E=\oplus_i^{k} \mathcal{E}_i$, then $E$ is polystable if
each $\mathcal{E}_i$ is a stable bundle with
$\mu(\mathcal{E}_1)=...=\mu(\mathcal{E}_k)=\mu({E})$
\cite{Donalson,UY01,UY02}. The Donaldson-Uhlenbeck-Yau theorem
\cite{Donalson,UY01,UY02} states that a (split) irreducible
holomorphic bundle $E$ admits a hermitian connection satisfying
Eq. (\ref{HYM}) if and only if $E$ is polystable.}. In other
words, one can either attempt to solve the Hermitian Yang-Mills
equations or, simply construct the associated stable holomorphic
vector bundles. It was shown in
\cite{Friedman:1997yq,Friedman:1997ma} that when $Z_3$ admits an
elliptic fibration, stable holomorphic bundles with structure
groups $SU(n)$ can be constructed by using spectral covers. In
what follows, we briefly review the spectral cover construction
and the computation of net chirality of the massless matter fields
in a four-dimensional effective
theory\cite{Friedman:1997yq,Donagi:2004ia,Other:global}.

Let $Z_3$ be an elliptically fibered Calabi-Yau threefold
$\pi_{H}:Z_{3}\ra B_2$ with a section $\sigma_{H}:B_2\ra Z_3$. Due
to the presence of the section $\sigma_H$, $Z_3$ can be described
by the Weierstrass model. One can realize $Z_3$ as a hypersurface
of $W\mathbb{P}_{2,3,1}^2$-fibration over $B_2$ given by
\begin{equation}
y^2=x^3+f_4xu^4+g_6u^6,\label{Z3}
\end{equation}
where $x$, $y$, $u$ are sections of $\mathcal{O}(2\sigma_H)\otimes
K_{B_2}^{-2}$, $\mathcal{O}(3\sigma_H)\otimes K_{B_2}^{-3}$, and
$\mathcal{O}(\sigma_H)$, respectively, while $f_4$ and $g_6$ are
sections of $K_{B_2}^{-4}$ and $K_{B_2}^{-6}$,
respectively.\footnote{The globally well-defined
$W\mathbb{P}_{2,3,1}^2$-fibration can be realized as the total
space of the weighted projective bundle $W\mathbb{P}(L^2\oplus
L^3\oplus \mathcal{O}_{B_2})$. It follows from the condition
$c_1(Z_3)=0$ that $L\cong K_{B_2}^{-1}$, where $K^{-1}_{B_2}$ is
the anticanonical bundle of $B_2$. Let
$c_1(\mathcal{O}_{P}(1))=\sigma_H$, then the homogeneous
coordinates $[x:y:u]$ are sections of
$\mathcal{O}(2\sigma_H)\otimes K_{B_2}^{-2}$,
$\mathcal{O}(3\sigma_H)\otimes K_{B_2}^{-3}$, and
$\mathcal{O}(\sigma_H)$, respectively.} Note that these sections
satisfy the following relation:
\begin{equation}
\sigma_H\cdot (\sigma_H+\pi_H^{\ast}c_1)=0,\label{sections
constraint Z_3}
\end{equation}
where $c_1\equiv c_1(B_2)$. At a generic point $b\in B_2$, the
fiber $\mathbb{E}_b$ is an elliptic curve. The restriction
$V|_{\mathbb{E}_b}$ of the bundle $V$ of rank $n$ to the elliptic
curve $\mathbb{E}_b$ is split. Namely, $V|_{\mathbb{E}_b}$ can be
decomposed as a direct summand of holomorphic line bundles. The
semi-stability of $V$ requires that these line bundles be all of
degree zero. Therefore, we can write
$V|_{\mathbb{E}_b}=\oplus_{i=1}^{n}\mathcal{O}_{\mathbb{E}_b}(q_i-e_0)$,
where $q_i\in\mathbb{E}_b $ and $e_0$ is a distinguished point
representing the identity element in the group law on
${\mathbb{E}_b}$. For $SU(n)$ bundles, it is required that
$c_1(V)=0$ which leads to the traceless condition $\sum_{i=1}^n
(q_i-e_0)=0$. When the point $b$ varies along $B_2$,
$\{q_1.q_2,...,q_n\}$ spans a $n$-fold cover over $B_2$, called
$SU(n)$ spectral cover. In particular, the $SU(4)$ spectral cover
is given by
\begin{equation}
\mathcal{C}^{(4)}_{H}:\; a_0u^4+a_2xu^2+a_3yu+a_4x^2=0,\label{cover
SU4}
\end{equation}
with a projection map
$p_{\mathcal{C}^{(4)}_{H}}:\mathcal{C}^{(4)}_H\ra B_2$. We denote
the homological class $[a_0]$ of the section $a_0$ by
$\pi_H^{\ast}\eta$, where $\eta\in H_2(B_2,\mathbb{Z})$ and write
the remaining sections as $[a_m]=\pi_H^{\ast}(\eta-mc_1)$, where
$m=2,3,4$.\footnote{Generically, the spectral cover defined by
Eq.~(\ref{cover SU4}) leads to a semistable
bundle\cite{Friedman:1997yq,Friedman:1997ma}. A sufficient
condition to obtain a holomorphic stable bundle $V$ is that
$\mathcal{C}^{(4)}_{H}$ is irreducible, which can be achieved by
imposing the following two conditions: $(1)$ The linear system
$|\eta|$ is base-point free in $B_2$, $(2)$ $\eta-mc_1$ is
effective in $B_2$\cite{Donagi:2004ia}. } The sections
$a_0,a_2,a_3$ and $a_4$ encode the information of deformation of
$\mathcal{C}^{(4)}_{H}$ defined by Eq.~(\ref{cover SU4}) and can
be regarded as complex moduli of the spectral cover. On the other
hand, the positions of the points $\{q_1,q_2,q_3,q_4\}$ or the
roots of the cover $\mathcal{C}^{(4)}_{H}$ characterize the
deformation of the bundle $V$. Therefore, $\{a_0,a_2,a_3,a_4\}$
characterize the deformation\footnote{The moduli space of stable
$SU(4)$ bundles on ${\mathbb{E}_b}$ is the projective space
$\mathbb{P}^3$. Fitting $\mathbb{P}^3$'s together, we obtain the
projective bundle $\mathbb{P}(\mathcal{O}_{B_2}\oplus
L^{-2}\otimes L^{-3}\oplus L^{-4})$ over $B_2$. In general, the
moduli space of stable $SU(n)$ bundles is the projective bundle
$\mathbb{P}(\mathcal{O}_{B_2}\oplus L^{-2}\otimes
L^{-3}\oplus...\oplus L^{-n})$\cite{Friedman:1997yq}. } of $V$. It
follows from Eq.~(\ref{cover SU4}) that the homological class of
$\mathcal{C}^{(4)}_H$ is given by
\begin{equation}
[\mathcal{C}^{(4)}_{H}]=4\sigma_H+\pi_H^{\ast}\eta.\label{class of
cover SU4}
\end{equation}
An $SU(4)$ bundle can be constructed by specifying a line bundle
or its twist $\gamma^{(4)}_H$ which is $(1,1)$-form on
$\mathcal{C}^{(4)}_H$. To obtain $SU(4)$ bundles, it is required
that $\gamma^{(4)}_H$ satisfies the traceless condition
$p_{\mathcal{C}^{(4)}_H\ast}\gamma^{(4)}_H=0$. This can be
achieved by setting
\begin{equation}
\gamma^{(4)}_H=(4-p_{\mathcal{C}^{(4)}_H}^{\ast}p_{\mathcal{C}^{(4)}_H\ast})([{\mathcal{C}}^{(4)}_H]\cdot\sigma_H).\label{gamma_H}
\end{equation}
Turning on an $SU(4)$ bundle over $Z_3$ breaks $E_8$ down to
$SO(10)$. Under the breaking pattern $E_8 ~\rightarrow~
SO(10)\times SU(4)$, the adjoint representation of $E_8$ is
decomposed as
\begin{equation}
\begin{array}{c@{}c@{}l@{}c@{}l}
E_8 &~\rightarrow~& SO(10)\times SU(4)\\
{\bf 248} &~\rightarrow~& ({\bf 1},{\bf 15})+({\bf 45},{\bf 1})+({\bf 10},{\bf 6})+({\bf 16},{\bf 4})+({\bf\overline{16}},{\bf\bar 4}).\\
\label{E_8 decomposition 01}
\end{array}
\end{equation}
The net chirality of matter fields can be calculated by the
Atiyah-Singer index theorem or by intersection numbers of matter
curves with
$\gamma_H^{(4)}$\cite{Friedman:1997yq,Donagi:2004ia,Other:global}.
Before computing the net chirality, we need to find the
homological classes of matter curves. The homological class of the
matter ${\bf 16}$ curve in $Z_3$ is given by the intersection of
$\mathcal{C}^{(4)}_{H}$ with the zero section
\begin{equation}
[\Sigma_{{\bf 16}, H}]=[\mathcal{C}^{(4)}_H]\cdot\sigma_H.
\end{equation}
The net chirality $N_{\bf 16}$ of the matter ${\bf 16}$ can be
evaluated by
\begin{eqnarray}
N_{\bf 16}&=&\int_{\Sigma_{{\bf 16},H}}
\gamma^{(4)}_{H}\nonumber\\&=& \gamma^{(4)}_H\cdot [\Sigma_{{\bf
16},H}]\nonumber\\&=&-\eta\cdot_{B_2}(\eta-4c_1).
\end{eqnarray}
To get the net chirality of matter ${\bf 10}$, we have to resolve
the singularity on the associated cover
$\mathcal{C}^{(6)}_{{\wedge^2 V},H}$ corresponding to the
antisymmetric representation ${\bf 6}$ in $SU(4)$. It can be done
by considering the intersection
$\mathcal{C}^{(4)}_H\cap\tau\mathcal{C}^{(4)}_H$, where $\tau$ is
a $\mathbb{Z}_2$ involution acting on the cover
$\mathcal{C}^{(4)}_H$ by $y\ra -y$ while keeping $x$ and $u$
untouched. More precisely, the intersection
$\mathcal{C}^{(4)}_H\cap\tau\mathcal{C}^{(4)}_H$ is determined by
\begin{equation}
\left\{\begin{array}{l} a_3yu=0\\
a_0u^4+a_2xu^2+a_4x^2=0.
\end{array}\label{matter 10 cover}   \right.
\end{equation}
The homological class of matter ${\bf 10}$ curve in $Z_3$ can be
computed as
\begin{eqnarray}
[\Sigma_{{\bf 10},
H}]&=&[\mathcal{C}^{(4)}_H]\cdot[\mathcal{C}^{(4)}_H]-[y]\cdot
[a_0u^4]-[u]\cdot [a_4x^2]\nonumber\\&=&
[\mathcal{C}^{(4)}_H]\cdot\{[\mathcal{C}^{(4)}_H]-3(\sigma_H+\pi_H^{\ast}c_1)-\sigma_H\}.
\end{eqnarray}
The net chirality $N_{\bf 10}$ can be calculated by the
intersection number\footnote{For the case of $SU(n)$ bundles,
$N_{\bf 16}=-\eta\cdot_{B_2}(\eta-nc_1)$ and $N_{\bf
10}=-(n-4)\eta\cdot_{B_2}(\eta-nc_1)$. The factor $(n-4)$ in
$N_{\bf 10}$ can be seen from the fact that $\chi(Z_3,\wedge^2
V)=(n-4)\chi(Z_3,V)$ where $Z_3$ is a Calabi-Yau threefold and $V$
is a vector bundle of rank $n$ with $c_1(V)=0$.} $ \gamma_H\cdot
[\Sigma_{{\bf 10},H}]$
\begin{eqnarray}
N_{\bf 10}&=& \gamma_H\cdot [\Sigma_{{\bf 10},H}]\nonumber\\&=&
[\mathcal{C}^{(4)}_H]\cdot
[4\sigma_H-\pi_H^{\ast}(\eta-4c_1)]\cdot\{[\mathcal{C}^{(4)}_H]-3(\sigma_H+\pi_H^{\ast}c_1)-\sigma_H\}
\nonumber\\&=&0.
\end{eqnarray}

\subsubsection{Del Pezzo Surface Construction}

In the previous section one can see that the information of the
bundle $V$ can be encoded in the spectral cover
$\mathcal{C}^{(4)}_H$ and the twist $\gamma^{(4)}_H$. However, the
construction can be translated to another form which involves del
Pezzo surfaces and is more suitable for the framework of
heterotic/F-theory duality. Before introducing the
heterotic/F-theory duality, we briefly review the del Pezzo
surface construction for $SU(4)$ bundles. Let $S$ be a del Pezzo
surface $dP_8$ which can be obtained by blowing up eight generic
points $p_1,p_2,...,p_8$ in $\mathbb{P}^2$. The second homology
group $H_2(S,\mathbb{Z})$ of $S$ is generated by the basis
$\{H,E_1,...,E_8\}$ with the intersection form given by
\begin{equation}
H\cdot H=1,\;\;\;\;\;H\cdot E_i=0,\;\;\;\;\;E_i\cdot
E_j=-\delta_{ij},\;\;\;i,j=1,2,...,8,
\end{equation}
where $H$ is the pullback of the hyperplane divisor in
$\mathbb{P}^2$ and $E_i$ are the exceptional divisors from
blow-ups. The anticanonical divisor $-K_{S}$ of $S$ is given by
\begin{equation}
-K_{S}=3H-\sum_{i=1}^8E_i.
\end{equation}
The linear system $|-K_{S}|$ has a base point and general elements
$\mathbb{E}$ of $|-K_{S}|$ are genus one curves.\footnote{Since
$-K_{S}$ is ample, $H^{0}(S,\mathcal{O}_{B_2}(-K_{S}))\neq 0$. The
linear system of $-K_{S}$ is defined by
$|-K_{S}|=\mathbb{P}H^{0}(S,\mathcal{O}_{S}(-K_{S}))$ and the base
point locus is defined by $\bigcap
\mathbb{E}_{\alpha},\;\mathbb{E}_{\alpha}\in |-K_{S}|$. For a del
Pezzo surface $dP_8$, it follows from the Riemann-Roch theorem and
Kodaira vanishing theorem that ${\rm dim}
(|-K_{S}|)=\chi(S,\mathcal{O}_{S}(-K_{S}))-1=(-K_{S})^2=1$. Since
$h^0(S,\mathcal{O}_S(-K_S))=2$, we have two homogeneous
polynomials of degree one and the base point is the unique common
zero $[0:0]$. Moreover, one can show that the linear system
$|-3K_S|$ induces a morphism $\Phi_{|-3K_S|}:S\ra
W\mathbb{P}^3_{2,3,1,1}$. The image of $\Phi_{|-3K_S|}$ is given
by Eq.~(\ref{dP_8 in WP2311})\cite{Other:global}. } Let us define
two subsects of $H_2(S,\mathbb{Z})$ as
follows\cite{Friedman:1997yq,Other:global}:
\begin{eqnarray}
&&I_8=\{l\in H_2(S,\mathbb{Z})|l\cdot l=-1,\;l\cdot (-K_{S})=1 \},  \\
&&R_8= \{C\in H_2(S,\mathbb{Z})|C\cdot C=-2,\;C\cdot(-K_{S})=0\}.
\end{eqnarray}
Note that $I_8$ and $R_8$ are in one-to-one correspondence through
$l=C+(-K_{S})$ and that the elements in $R_8$ are in one-to-one
correspondence with roots of $E_8$. The generators of $R_8$ can be
chosen as follows:
\begin{equation}
C_k=E_k-E_{k+1},\;k=1,2,...,7,\;\;\;\;\;C_8=H-(E_6+E_7+E_8).
\end{equation}
The intersection matrix of $R_8$ is given by $(-C_{E_8})$ where
$C_{E_8}$ is the Cartan matrix of $E_8$. Given $\mathbb{E}\in
|-K_{S}|$, a flat bundle on $\mathbb{E}$ is given by
\begin{equation}
\mathcal{O}_{\mathbb{E}}(C_k|_{\mathbb{E}})\cong
\mathcal{O}_{\mathbb{E}}(q_k-e_0),\label{flat bundle in dP_8}
\end{equation}
where $C_k\in R_8$ and $q_k\in \mathbb{E}$ given by $l_k\cdot
(-K_{S})$. Recall that the spectral cover describes a flat bundle
on an elliptic fiberation $\pi_H:Z_3\ra B_2$ by specifying a set
$\{q_k\}$ for each fiber $\mathbb{E}_b,\;b\in B_2$. Equivalently,
one can describe the bundle by starting with embedding an elliptic
curve $\mathbb{E}_b$ into a fiber of $dP_8$-fibration over $B_2$
$\pi_{W_4}: W_{4}\ra B_2$ with ${\pi_{W_4}}|_{Z_3}=\pi_H$. Then
the local data $V|_{\mathbb{E}_b}$ of the bundle $V$ can be
described by the cycles $\{C_1,C_2,...,C_n\}$ in $R_8$ via
Eq.~(\ref{flat bundle in dP_8}). On the other hand, one can
realize a $dP_8$ surface as a divisor in
$W\mathbb{P}^3_{2,3,1,1}$. More precisely, a $dP_8$ surface in
$W\mathbb{P}^3_{2,3,1,1}$ can be described by the Weierstrass
model as follows:
\begin{equation}
y^2=x^3+\tilde{f}_4(Z_1,Z_2)x+\tilde{g}_6(Z_1,Z_2),\label{dP_8 in
WP2311}
\end{equation}
where $[x:y:Z_1:Z_2]$ are homogeneous coordinates of
$W\mathbb{P}^3_{2,3,1,1}$, $\tilde{f}_4$ and $\tilde{g}_6$ are
homogeneous polynomials of degree four and six, respectively.
Through this embedding, one can find that the bundle moduli of a
flat bundle on $\mathbb{E}$ map to the complex structure moduli of
the defining equation Eq.~(\ref{dP_8 in WP2311}). For the case of
$G=SU(4)$, one can construct the bundle through the spectral cover
construction by specifying points $\{q_1,q_2,q_3,q_4\}$ on
$\mathbb{E}_b$. The bundle moduli are characterized by the
coefficients $\{a_0,a_2,a_3,a_4\}$ of the spectral cover defined
by Eq.~(\ref{cover SU4}). Equivalently, this data can be described
by the (-2)-cycles $\{C_1,C_2,C_3,C_4\}$ in $dP_8$ and their
intersection numbers. The intersection of these cycles form the
Cartan matrix of $SU(4)$. The complement of the extended Dynkin
diagram of $SU(4)$ in $E_8$ corresponds to the vanishing cycles
which leads to a $D_5$ singularity in $dP_8$. In other words, the
unbroken GUT group $SO(10)$ corresponds to a $D_5$ singularity in
$dP_8$. Therefore, one can construct an $SO(10)$ GUT group by
engineering a $D_5$ singularity in $dP_8$. More precisely, one can
consider the Weierstrass model
\begin{equation}
y^2=x^3+f_4Z_1^4x+g_6Z_1^6+Z_2Z_1(b_0Z_1^4+b_2Z_1^{2}x+b_3Z_1y+b_4x^2).\label{resolved
dP_8 in WP2311}
\end{equation}
Note that $Z_2=0$ locus is an elliptic curve given by the
Weierstrass equation $y^2=x^3+f_4x+g_6$ and that the parenthesis
in Eq.~(\ref{resolved dP_8 in WP2311}) reduces to the spectral
cover $\mathcal{C}_H^{(4)}$ given by Eq.~(\ref{cover SU4}) when
$Z_1\ra u$ with $b_m|_{Z_3}=a_m$. It is clear that in this case
the bundle moduli $\{a_0,a_2,a_3,a_4\}$ map to the complex moduli
of $dP_8$ given by Eq.~(\ref{resolved dP_8 in WP2311}). The dual
F-theory geometry can be described as a $dP_9$-fibration over
$B_2$, which is obtained by blowing up the base point. The $dP_8$
construction described above for $SU(n)$ bundles can be realized
by a $dP_9$ surface whose intersection matrix of $(-2)$-cycles
contains the Cartan matrix of $E_8$ . It can be seen by
taking\cite{Marsano:2010ix,Other:global}
\begin{eqnarray}
&&I_8=\{l\in H_2(dP_9,\mathbb{Z})|l\cdot l=-1,\;l\cdot (-K_{dP_9})=1,\;l\cdot E_9=0 \},  \\
&&R_8=\{C\in H_2(dP_9,\mathbb{Z})|C\cdot
C=-2,\;C\cdot(-K_{dP_9})=0,\;C\cdot E_9=0\},
\end{eqnarray}
where $E_9$ is an exceptional divisor from the blow-up of the base
point. The geometry of a $dP_9$-fibration can be obtained by
taking the stable degenerate limit of a $K3$-fibration on $B_2$ in
F-theory\cite{Friedman:1997yq,Aspinwall:1997eh,Donagi:1998vw,Donagi:2008ll,Other:global}.
Through this degenerate limit, we can embed the data of the bundle
$V$ into dual F-theory geometry. We shall describe this degenerate
limit in the next section.

\subsection{Heterotic/F-theory Duality}

\subsubsection{Stable Degeneration Limit}

Let us consider F-theory on an elliptically fibered Calabi-Yau
fourfold $\pi_{X_4}:X_4\ra B_3$ with a section
$\sigma_{B_3}:B_3\ra X_4$. With the section $\sigma_{F}$, $X_4$
can be described by the Weierstrass model:
\begin{equation}
y^2=x^3+fxu^4+gu^6.\label{Weierstrass model}
\end{equation}
The Calabi-Yau condition $c_1(X_4)=0$ requires that $f$ and $g$
are sections of $K_{B_3}^{-4}$ and $K_{B_3}^{-6}$,
respectively.\footnote{To see this, we can embed $X_4$ as a
section of a weighted projective bundle over $B_3$. More
precisely, we homogenize Eq.~$(\ref{Weierstrass model})$ to be
$y^2=x^3+xu^4+gu^6\hookrightarrow W\mathbb{P}^2_{2,3,1}$, where
$f$ and $g$ are sections of line bundles $L^4$ and $L^4$ on $B_3$,
respectively. To obtain a globally well-defined fibration, let
${\bar X}_5$ be the total space of the weighted projective bundle
$W\mathbb{P}(L^2\otimes L^3\otimes \mathcal{O}_{B_3})$ over $B_3$
and consider $X_4$ to be a hypersurface in ${\bar X}_5$. By the
adjuction formula\cite{Hartshone:01,Griffith:01}, we have
$c(X_4)=\frac{c(B_3)(1+2r+2\pi^{\ast}_{X_4}t)(1+3r+3\pi^{\ast}_{X_4}t)(1+r)}{(1+6r+6\pi^{\ast}_{X_4}t)}$,
where $r\equiv c_1(\mathcal{O}_P(1))$ and $t\equiv c_1(L)$. It
follows from the condition $c_1(X_4)=0$ that $L=K_{B_3}^{-1}$.}
The heterotic/F-theory duality requires that $B_3$ admits a
$\mathbb{P}^{1}$-fibration over some surface $B_2$. Let
$[Z_1:Z_2]$ be the homogeneous coordinates of $\mathbb{P}^1$
fiber. Since $f$ and $g$ are the homogeneous polynomials of degree
$8$ and $12$ in terms of $[Z_1:Z_2]$, respectively,
\footnote{Recall that the anticanonical bundle
$K^{-1}_{\mathbb{P}^{n}}$ of $n$-dimensional complex projective
space $\mathbb{P}^n$ is $K_{\mathbb{P}^{n}}^{-1}=(n+1)H\equiv
\mathcal{O}_{\mathbb{P}^n}(n+1)$. So
$K^{-4}_{\mathbb{P}^1}=\mathcal{O}_{\mathbb{P}^1}(8)$ and
$K^{-6}_{\mathbb{P}^1}=\mathcal{O}_{\mathbb{P}^1}(12)$.} one can
expand Eq.~(\ref{Weierstrass model}) as
\begin{equation}
y^2=x^3+\big(\sum_{i=0}^{8}f_iZ_1^iZ_2^{8-i}\big)xu^4+\big(\sum_{j=0}^{12}g_jZ_1^{j}Z_2^{12-j}\big)u^6,\label{Weierstrass
model02}
\end{equation}
where $f_i$ and $g_j$ are sections of suitable line bundles over
$B_2$. When $Z_1\ra 0$ and set $Z_2=1$, Eq.~(\ref{Weierstrass
model02}) becomes\cite{Donagi:1998vw,Other:global,Marsano:2010ix}
\begin{equation}
y^2=x^3+\big(\sum_{i=0}^{4}f_iz_1^i\big)xu^4+\big(\sum_{j=0}^{6}g_jz_1^{j}\big)u^6,\label{Weierstrass
model degeneration01}
\end{equation}
where $z_1\equiv \frac{Z_1}{Z_2}$. On the other hand, taking
$Z_2\ra 0$ and set $Z_1=1$, Eq.~(\ref{Weierstrass model02})
becomes
\begin{equation}
y^2= x^3+\big(\sum_{m=0}^{4}f_{m+4}z_2^{4-m}\big)xu^4
+\big(\sum_{l=0}^{6}g_{l+6}z_2^{6-l}\big)u^6,\label{Weierstrass
model degeneration02}
\end{equation}
where $z_2\equiv \frac{Z_2}{Z_1}$. These two limits correspond to
two $dP_9$ surfaces\footnote{The hypersurfaces described by
Eq.~(\ref{Weierstrass model degeneration01}) and
Eq.~(\ref{Weierstrass model degeneration02}) both are homogeneous
polynomials of degree six in $W\mathbb{P}^{3}_{2,3,1,1}$. They are
actually $dP_8$ surfaces. It follows from the adjuction formula
that $c_1(S)=x$ and $c_2(S)=11x^2$, where $x\equiv r+t$. By the
Riemann-Roch theorem $12\chi(\mathcal{O}_S)=c_1^2(S)+c_2(S)$, we
obtain $x^2=1$ and then Euler characteristic $\chi(S)=11$. For
$dP_k$ surfaces, $\chi(dP_k)=3+k$, which implies that $k=8$. One
can obtain $dP_9$'s by blowing up the point $Z_1=Z_2=0$.} glued
together along an elliptic curve $\mathbb{E}$ with the Weierstrass
equation\footnote{The elliptic curve $\mathbb{E}$ is an effective
divisor of the linear system $|-K_{S}|$. By the adjunction
formula, we obtain $2g-2=\mathbb{E}(\mathbb{E}+K_{S})=0$, which
implies that $\mathbb{E}$ is an elliptic curve.}:
\begin{equation}
y^2=x^3+f_4xu^4+g_6u^6\label{Elliptic curve E}.
\end{equation}
This elliptically fibered Calabi-Yau threefold $\pi_{H}:Z_3\ra
B_2$ is the background of heterotic string compactification. Two
$dP_9$ surfaces, Eq.~$(\ref{Weierstrass model degeneration01})$
and $(\ref{Weierstrass model degeneration02})$ encode the data of
bundles $E_8\times E_8$ in the heterotic string. With
heterotic/F-theory duality, one can find that constructing an
stable $SU(4)$ bundle on an elliptically fibered $Z_3$ with a base
$B_2$ by using spectral cover construction corresponds to
engineering an $D_5$ singularity in the geometry of
$dP_9$-fibration on $B_2$ given by Eq.~(\ref{resolved dP_8 in
WP2311}).

\subsubsection{Dual $SU(4)$ spectral Divisor in F-theory}

Let $Y_4$ be a $dP_9$-fibration over a complex surface $B_2$ with
a projection map $p: Y_4\ra B_2$. Since $dP_9$ is an elliptic
surface, $Y_4$ can be regarded as an elliptic fibration over a
threefold $B_3$ with a section $\sigma_{F}: B_3\ra Y_4$ and $B_3 $
admits a $\mathbb{P}^1$-fibration over $B_2$. The projection map
of the elliptic fibration and $\mathbb{P}^1$-fibration are denoted
by $\pi_{F}:Y_4\ra B_3$ and $\va:B_3\ra B_2$, respectively. To
describe $Y_4$, we embed the elliptic fiber as a divisor of
$W\mathbb{P}^2_{2,3,1}$ with homogeneous coordinates $[x:y:u]$ and
consider the following Weierstrass model:
\begin{eqnarray}
y^2&=&x^3+f_4(Z_1u)^4x+g_6(Z_1u)^6+Z_2(Z_1u)^{5-n}[b_0(Z_1u)^n+b_2(Z_1u)^{n-2}x
\nonumber\\&+&b_3(Z_1u)^{n-3}y+...],\label{dP_8 fibration in
WP2311}
\end{eqnarray}
where the last term in the bracket is $b_nx^{n/2}$ for $n$ even,
or $b_nx^{(n-3)/2}y$ for $n$ odd. Note that $x$, $y$, and $u$ are
sections of $\mathcal{L}^2$, $\mathcal{L}^3$, and
$\mathcal{O}_{B_3}$, respectively, where $\mathcal{L}$ is a line
bundle on $B_3$ and will be determined later. To make
$W\mathbb{P}^2_{2,3,1}$-fibration globally well-defined, we
consider $Y_4$ be a divisor in the weighted projective bundle
$W\mathbb{P}(\mathcal{L}^2\oplus\mathcal{L}^3\oplus\mathcal{O}_{B_3})$.
We denote the fiber by $\mathcal{O}_P(1)$. Let
$c_1(\mathcal{O}_P(1))=\sigma_F$ and $c_1(\mathcal{L})=l$. By
using the adjuction formula, we obtain
\begin{equation}
c(Y_4)=c(B_3)\frac{(1+2\sigma_F+2\pi^{\ast}_F
l)(1+3\sigma_F+3\pi^{\ast}_F
l)(1+\sigma_F)}{(1+6\sigma_F+6\pi^{\ast}_F l)},\label{adjuction eq
01}
\end{equation}
where $c$ stand for the total Chern class. It follows from
Eq.~(\ref{adjuction eq 01}) that
\begin{equation}
c_1(Y_4)=\pi^{\ast}_F c_1(B_3)-\pi^{\ast}_F l.\label{c_1 for Y_4}
\end{equation}
Let us turn to the geometry of $B_3$. We take $B_3$ to be a
$\mathbb{P}^1$ bundle over $B_2$. To be concrete, let
$B_3=\mathbb{P}(\mathcal{O}_{B_2}\oplus\mathcal{M})$ with
$c_1(\mathcal{O}_P(1))=r$ and $c_1(\mathcal{M})=t$, where
$\mathcal{M}$ is a line bundle on $B_2$. By using the adjuction
formula, we have
\begin{equation}
c(B_3)=c(B_2)(1+r)(1+r+\varphi^{\ast}t),\label{adjuction eq 02}
\end{equation}
which implies that
\begin{equation}
\left\{\begin{array}{l} c_1(B_3)=2r+\va^{\ast}(c_1+t)\\
c_2(B_3)=\va^{\ast} c_2+\va^{\ast}c_1\cdot(\va^{\ast}t+2r)\\
c_3(B_3)=\va^{\ast} c_2\cdot(\va^{\ast}t+2r)
\end{array}\label{B_3 Chern class},   \right.
\end{equation}
where $c_1=c_1(B_2)$ and the relation $r\cdot (r+\va^{\ast}t)=0$
has been used. On the other hand, it follows from Eq.~(\ref{dP_8
fibration in WP2311}) that the heterotic Calabi-Yau threefold
$Z_3$ is given by $Z_2=0$ which is a submanifold of $Y_4$. By
using the adjuction formula, we have
\begin{equation}
c(Z_3)=\frac{c(Y_4)}{(1+\pi_F^{\ast}r+p^{\ast}t)}.\label{adjuction
eq 03}
\end{equation}
It follows from the Calabi-Yau condition $c_1(Z_3)=0$,
Eq.~(\ref{c_1 for Y_4}), and Eq.~(\ref{B_3 Chern class}) that
\begin{equation}
c_1(Y_4)=\pi_F^{\ast}r+p^{\ast}t,\;\;\; \pi^{\ast}_F
l=\pi^{\ast}_F
c_1(B_3)-\pi_F^{\ast}r-p^{\ast}t=\pi^{\ast}_Fr+p^{\ast}c_1.\label{adjuction
eq 03}
\end{equation}
Therefore, the homological classes of sections appearing in
Eq.~(\ref{except lines SU4}) are as follows:
\begin{equation}
[x]=2(\sigma_F+\pi^{\ast}_F
r+p^{\ast}c_1),\;\;[y]=3(\sigma_F+\pi^{\ast}_F
r+p^{\ast}c_1),\;\;[u]=\sigma_F,\label{sections in dual X_4 01}
\end{equation}
\begin{equation}
[Z_1]=\pi^{\ast}_F r,\;\;[Z_2]=\pi^{\ast}_F
r+p^{\ast}t,\;[b_m]=p^{\ast}[(6-m)c_1-t],\;\;m=0,2,3,4.\label{sections
in dual X_4 02}
\end{equation}
Following the proposal in\cite{Marsano:2010ix}, we define the
spectral divisor $\mathcal{D}_F^{(n)}$ of $Y_4$ by
\begin{equation}
\mathcal{D}_F^{(n)}:b_0(Z_1u)^n+b_2(Z_1u)^{n-2}x+b_3(Z_1u)^{n-3}y+...=0,\label{Spectral
divisor}
\end{equation}
where the last term is $b_nx^{n/2}$ for $n$ even, or
$b_nx^{(n-3)/2}y$ for $n$ odd. The projection map is denoted by
$p_{\mathcal{D}_F^{(n)}}:\mathcal{D}_F^{(n)}\ra B_3$. Let
$\gamma_F^{(n)}$ be a $(1,1)$ form on $\mathcal{D}_F^{(n)}$. It
was proposed in\cite{Marsano:2010ix} that the net chirality
formula for matter in the representation $\mathbf{r}$ can be
computed as
\begin{equation}
N_{\mathbf{r}}=[\widehat{\Sigma}_{\mathbf{r}}]\cdot\mathcal{G}^{(n)}_{F}\cdot
p^{\ast}_{\mathcal{D}_F^{(n)}} B_2,\label{Net Chiralty in
F-theory}
\end{equation}
where $[\widehat{\Sigma}_{\mathbf{r}}]$ is the dual matter surface
inside $\mathcal{D}_F^{(n)}$ and $\mathcal{G}^{(n)}_{F}$ is
defined by $\gamma_F^{(n)}=[\mathcal{D}_F^{(n)}]\cdot
\mathcal{G}^{(n)}_{F}$ for given $\gamma_F^{(n)}$. For the case of
$n=4$, we have
\begin{equation}
y^2=x^3+f_4(Z_1u)^4x+g_6(Z_1u)^6+Z_2[b_0(Z_1u)^5+b_2(Z_1u)^3x+b_3(Z_1u)^2y+b_4(Z_1u)x^2].\label{dP9}
\end{equation}
Note that when $Z_2=0$, Eq.~(\ref{dP9}) reduces to $Z_3$ defined
by Eq.~(\ref{Z3}). In this case the spectral divisor is given by
\begin{equation}
\mathcal{D}^{(4)}_{F}:\;b_0(Z_1u)^4+b_2(Z_1u)^2x+b_3(Z_1u)y+b_4x^2=0,\label{except
lines SU4}
\end{equation}
with a projection map $p_{\mathcal{D}^{(4)}_F}:
\mathcal{D}^{(4)}_F\ra B_3$. The divisor $\mathcal{D}^{(4)}_{F}$
can be realized as the union of four exceptional lines of $dP_9$
comprising a fundamental representation of
$SU(4)$\cite{Donagi:2008ll,Curio:1998bva,Other:global}. With
Eq~(\ref{sections in dual X_4 01}) and Eq.~(\ref{sections in dual
X_4 02}), the homological class of $\mathcal{D}^{(4)}_F$ is given
by
\begin{equation}
[\mathcal{D}^{(4)}_F]=4(\sigma_F+\pi_F^{\ast}r)+p^{\ast}(6c_1-t).
\end{equation}
The traceless flux $\gamma^{(4)}_F$ can be computed as
\begin{eqnarray}
\gamma^{(4)}_F&=&(4-p^{\ast}_{\mathcal{D}^{(4)}_F}p_{\mathcal{D}^{(4)}_F\ast})([\mathcal{D}^{(4)}_F]\cdot\sigma_F)\nonumber\\
&=&[\mathcal{D}^{(4)}_F]\cdot[4\sigma_F-p^{\ast}(2c_1-t)],\label{dual
twist}
\end{eqnarray}
where the relation
$\sigma_F\cdot(\sigma_F+\pi_F^{\ast}r+p^{\ast}c_1)=0$ has been
used. It follows from Eq.~(\ref{dual twist}) that
$\mathcal{G}^{(4)}_F=4\sigma_F-p^{\ast}(2c_1-t)$. To calculate the
chiral spectrum, we need to calculate the homological classes of
dual matter surfaces. The dual matter surface
$\widehat{\Sigma}_{\bf 16}$ sits in the locus of the intersection
$\{(Z_1u)=0\}\cap\{b_4=0\}$ and then its homological class is
given by
\begin{equation}
[\widehat{\Sigma}_{\bf 16}]=(\sigma_F+\pi^{\ast}_F r)\cdot
p^{\ast}(2c_1-t).
\end{equation}
By using the net chirality formula Eq.~(\ref{Net Chiralty in
F-theory}), we obtain
\begin{eqnarray}
N_{\bf 16}&=&[\widehat{\Sigma}_{\bf
16}]\cdot\mathcal{G}^{(4)}_F\cdot\pi_F^{\ast}r\nonumber\\&=&-(6c_1-t)\cdot_{B_2}(2c_1-t).
\end{eqnarray}
On the other hand, the dual matter surface $\widehat{\Sigma}_{\bf
10}$ sits in the locus of
$\mathcal{D}^{(4)}_F\cap\tau\mathcal{D}^{(4)}_F$ where $\tau$ is a
$\mathbb{Z}_2$ involution $y\ra -y$ acting on
$\mathcal{D}^{(4)}_F$ while keeping $x$, $u$, and $Z_1$ intact.
More precisely, the intersection loci of
$\mathcal{D}^{(4)}_F\cap\tau\mathcal{D}^{(4)}_F$ are determined by
\begin{equation}
\left\{\begin{array}{l} b_3(Z_1u)y=0\\
b_0(Z_1u)^4+b_2(Z_1u)^2x+b_4x^2=0.
\end{array}\label{matter 10 surface}   \right.
\end{equation}
It follows from Eq.~(\ref{matter 10 surface}) that the homological
class of dual matter surface $\widehat{\Sigma}_{\bf 10}$ is
\begin{eqnarray}
[\widehat{\Sigma}_{\bf
10}]&=&[\mathcal{D}^{(4)}_F]\cdot[\mathcal{D}^{(4)}_F]-[Z_1u]\cdot[b_4]-[y][b_4x^2]-2[x][Z_1]\nonumber\\&=&
(\sigma_F+\pi_F^{\ast}r)\cdot
p^{\ast}(12c_1-4t)+p^{\ast}(6c_1-t)\cdot p^{\ast}(3c_1-t).
\end{eqnarray}
By using Eq.~(\ref{Net Chiralty in F-theory}), the net chirality
of matter ${\bf 10}$ is
\begin{eqnarray}
N_{\bf 10}&=&[\widehat{\Sigma}_{\bf
10}]\cdot\mathcal{G}^{(4)}_F\cdot\pi_F^{\ast}r\nonumber\\&=&0.
\end{eqnarray}
These results agree with the computations in the dual heterotic
string framework by identifying
${\mathcal{D}}^{(4)}_{F}|_{Z_3}=\mathcal{C}^{(4)}_H$ and
$b_m|_{Z_3}=a_m$, which gives rise to the relation $\eta=6c_1-t$.
It was argued in\cite{Marsano:2010ix} that the chirality formula
Eq.~(\ref{Net Chiralty in F-theory}) can be applied to the cases of
F-theory compactifications without heterotic duals. In section 3, we shall
briefly review semi-local $SU(4)$ cover construction
\cite{Chen:2010tp} and its global completion\cite{Marsano:2010ix}.

\section{Global Completion of $SU(4)$ Cover}

In this section we shall discuss the case of an F-theory
compactification on an elliptically fibered Calabi-Yau fourfold
without a heterotic dual. We first briefly review the semi-local
$SU(4)$ spectral cover construction studied in\cite{Chen:2010tp}.
In the second part we construct the $SU(4)$ spectral divisor
following the proposal in \cite{Marsano:2010ix}.

\subsection{Semi-local $SU(4)$ Cover}

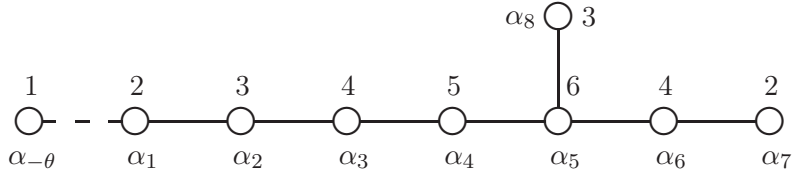
\begin{figure}[h]
\center \large
\begin{picture}(400,80)
\thicklines
\multiput(60, 20)(40, 0){8}{\circle{10}} %
\put(260, 60){\circle{10}} \put(260, 25){\line(0,1){30}}%
\multiput(105, 20)(40, 0){6}{\line(1, 0){30}} %
\dashline[4]{6}(65,20)(95,20) %
\put(58,30){\small 1} \put(98,30){\small 2} \put(138,30){\small 3}
\put(178,30){\small 4} \put(218,30){\small 5} \put(263,30){\small
6} \put(298,30){\small 4} \put(338,30){\small 2}
\put(269,56){\small 3} %
\put(52,4){\small $\alpha_{-\theta}$} \put(97,4){\small
$\alpha_1$} \put(137,4){\small $\alpha_2$} \put(177,4){\small
$\alpha_3$} \put(217,4){\small $\alpha_4$} \put(257,4){\small
$\alpha_5$} \put(297,4){\small $\alpha_6$} \put(337,4){\small
$\alpha_7$} \put(240,57){\small $\alpha_8$}
\end{picture}
\caption{The extended $E_8$ Dynkin diagram and
indices}\label{Dynkin}
\end{figure}

Let us consider an elliptically fibered Calabi-Yau fourfold $\pi:
Z_4\ra B_3$ with a section $\sigma:B_3\ra Z_4$ and $B_2$ to be a
divisor in $B_3$ where $Z_4$ exhibits a $D_5$ singularity.
Generically, $Z_4$ can be described by the Tate form as follows:
\begin{equation}
y^2=x^3+b_4x^2z+b_3yz^2+b_2xz^3+b_0z^5.\label{Tate form}
\end{equation}
Let us define $t\equiv -c_1(N_{B_2/B_{3}})$ and then the
homological classes of the sections $x$, $y$, $z$, and $b_m$ can
be expressed as
\begin{equation}
[x]=3(c_1-t),\;[y]=2(c_1-t),\;[z]=-t,\; [b_m]=
(6-m)c_1-t=\eta-mc_1.
\end{equation}
Recall that locally $Z_4$ can be described by an ALE fibration
over $B_2$. Pick a point $p\in B_2$, the fiber is an ALE space
denoted by ${\rm ALE}_p$. The ALE space can be constructed by
resolving an orbifold $\mathbb{C}^2/\Gamma_{ADE}$, where
$\Gamma_{ADE}$ is a discrete subgroup of $SU(2)$.\footnote{For
more information, see \cite{Douglas:1996sw, Math3, Math2, Math1,
Math4, Math5}.} It was shown that the intersection matrix of the
exceptional 2-cycles corresponds to the Cartan matrix of $ADE$
type, which can be described by $ADE$ Dynkin diagrams. Let us take
$\alpha_i\in H_{2}({\rm ALE}_p,\mathbb{Z}),\: i=1,2,...,8$ to be
the roots\footnote{By abuse of notation, the corresponding
exceptional 2-cycles are also denoted by $\alpha_i$.} of $E_8$ and
the extended $E_8$ Dynkin diagram with roots and Dynkin indices to
be shown in Fig \ref{Dynkin}.
Notice that $\alpha_{-\theta}$ is the highest root and satisfies
the condition
$\alpha_{-\theta}+2\alpha_1+3\alpha_2+4\alpha_3+5\alpha_4+6\alpha_5+4\alpha_6+2\alpha_7+3\alpha_8=0$.
To obtain $SO(10)$, we take the volume of the cycles
$\{\alpha_4,\alpha_5,...,\alpha_8\}$ to be vanishing and then
$SU(4)$ is generated by $\{\alpha_1,\alpha_2,\alpha_3\}$. An
enhancement to $E_6$ happens when $\alpha_3$ or any of its images
under the Weyl permutation shrinks to zero size. We define
$\{\lambda_1,...,\lambda_4\}$ to be the periods 
of these cycles.
As described in \cite{Donagi:2008sl, Donagi:global}, theses
$\lambda_i$ are encoded in the coefficients $b_m$ as follows:
\begin{equation}
\left\{\begin{array}{l}\displaystyle \sum_{i}\lambda_i=\frac{b_1}{b_0}=0\\
\displaystyle\sum_{i<j}\lambda_i\lambda_j=\frac{b_2}{b_0}\\
\displaystyle\sum_{i<j<k}\lambda_i\lambda_j\lambda_k=\frac{b_3}{b_0}\\
\displaystyle\prod_{l}\lambda_l=\frac{b_4}{b_0}.
\end{array}\label{Lambda}   \right.
\end{equation}
Equivalently, $\{\lambda_1,...,\lambda_4\}$ are the roots of the
equation
\begin{equation}
b_0\prod_{k=1}^4 (s+\lambda_k)=b_0s^4+b_2s^2+b_3s+b_4=0.
\label{affine SU(4) cover}
\end{equation}
When $p\in B_2$ varies along $B_2$, Eq.~(\ref{affine SU(4) cover})
defines a fourfold cover $\mathcal{C}^{(4)}$ over $B_2$, the
semi-local $SU(4)$ spectral cover. This cover can be described as
a section of the canonical bundle $K_{B_2}\ra B_2$. When
$\lambda_i$ vanish, $\prod_{i}\lambda_i=b_4=0$ and the
corresponding gauge group is enhanced to $E_6$, which implies that
the matter field ${\bf 16}$ is localized at the locus $\{b_4=0\}$.
On the other hand,the matter field ${\bf 10}$ corresponds to the
anti-symmetric representation ${\bf 6}$ of $SU(4)$, associated
with a sixfold cover $\mathcal{C}_{\wedge^2 V}^{(6)}$ over $B_2$.
This associated cover $\mathcal{C}^{(6)}_{\wedge^2 V}$ is given by
\begin{equation}
\mathcal{C}^{(6)}_{\wedge^2 V}:\;\; b_0^2\prod_{i<j} (s+\lambda_i
+\lambda_j)=b_0^2 s^6+ 2b_0 b_2 s^4 +( b_2^2 -4b_0 b_4)s^2 -
b_3^2=0.\label{antisymmetry cover}
\end{equation}
Since matter ${\bf 10}$ corresponds to
$\lambda_i+\lambda_j=0,\;i\neq j$, it follows from
Eq.~(\ref{antisymmetry cover}) that $b_3=0$, which means that
matter ${\bf 10}$ is localized at the locus $\{b_3=0\}$ as we
expected from the $D_6$ singularity of Eq.~(\ref{Tate form}). From
the discussion above, we see that spectral cover indeed encodes
the information of singularities and gauge group enhancements.
Moreover, we can construct a Higgs bundle to calculate the chiral
spectrum for matter ${\bf 16}$ and ${\bf 10}$ by switching on a
line bundle on the cover.
Let us define $X$ to be the total space
of the canonical bundle $K_{B_2}$ over $B_2$. Note that $X$ is a
local Calabi-Yau threefold.
but $X$ is non-compact
. To obtain a compact space, one can compactify $X$ to the total
space $\bar X$ of the projective bundle over $B_2$, $i.e.$
\begin{equation}
\bar{X}= {\mathbb{P}}(\mathcal{O}_{B_2}\oplus K_{B_2}),
\end{equation}
with a projection map $\pi:{\bar X}\ra B_2$, where
$\mathcal{O}_{B_2}$ is the trivial bundle over $B_2$. Notice that
$\bar X$ is compact but no longer a Calabi-Yau threefold. Let
$\mathcal{O}_{P}(1)$ be a hyperplane section of ${\mathbb{P}}^1$
fiber and denote its first Chern class by $\sigma_{\infty}$. We
define the homogeneous coordinates of the fiber by $[U:W]$. Note
that $\{U=0\}$ and $\{W=0\}$ are sections of
$\mathcal{O}_P(1)\otimes K_S$ and $\mathcal{O}_P(1)$, while the
class of $\{U=0\}$ and $\{W=0\}$ are
$\sigma\equiv\sigma_{\infty}-\pi^{\ast}c_1$ and $\sigma_{\infty}$,
respectively. By the emptiness of intersection of $\{U=0\}$ and
$\{W=0\}$, we obtain
$\sigma\cdot\sigma=-\sigma\cdot\pi^{\ast}c_1$. We define the
affine coordinate $s$ by $s=U/W$ and then the $SU(4)$ cover given
by Eq.~(\ref{affine SU(4) cover}) can be written as
\begin{equation}
\mathcal{C}^{(4)}:\;\;\;b_0U^4+b_2U^2W^2+b_3UW^3+b_4W^4=0\label{homo
SU(4) cover}
\end{equation}
with a projection map $p_{\mathcal{C}^{(4)}}: \mathcal{C}^{(4)}\ra
B_2$. It is not difficult to see that the homological class
$[\mathcal{C}^{(4)}]$ of the cover $\mathcal{C}^{(4)}$ is given by
$[\mathcal{C}^{(4)}]=4\sigma+\pi^{\ast}\eta$. We can calculate the
matter ${\bf 16}$ curve by intersecting $[\mathcal{C}^{(4)}]$ with
$\sigma$
\begin{equation}
[\mathcal{C}^{(4)}]\cap\sigma=(4\sigma+\pi^{\ast}\eta)\cdot\sigma=\sigma\cdot\pi^{\ast}(\eta-4c_1).
\end{equation}
On the other hand, it follows from Eq.~(\ref{antisymmetry cover})
that the homological class of the cover
$\mathcal{C}^{(6)}_{\wedge^2V}$ is given by
\begin{equation}
[\mathcal{C}^{(6)}_{\wedge^2V}] =6\sigma+2\pi^{\ast}\eta.
\end{equation}
However, the cover $\mathcal{C}^{(6)}_{\wedge^2 V}$ is generically
singular. To solve this problem, one can consider intersection
$\mathcal{C}^{(4)} \cap \tau \mathcal{C}^{(4)}$ and define
\cite{Donagi:2004ia}
\begin{eqnarray}
[D]=[\mathcal{C}^{(4)}]\cap ([\mathcal{C}^{(4)}]
-3\sigma_{\infty}-\sigma).\label{D curve}
\end{eqnarray}
where $\tau$ is a $\mathbb{Z}_2$ involution $W\ra -W$ acting on
the spectral cover $\mathcal{C}^{(4)}$. To obtain chiral spectrum,
we turn on a spectral line bundle $\mathcal{L}$ on the cover
$\mathcal{C}^{(4)}$. The corresponding Higgs bundle is given by
$p_{\mathcal{C}^{(4)}\ast}\mathcal{L}$. For $SU(n)$ bundles, it is
required that $c_1(p_{\mathcal{C}^{(4)}\ast}\mathcal{L})=0$. It
follows that
\begin{equation}
p_{\mathcal{C}^{(4)}\ast}c_1(\mathcal{L})-\frac{1}{2}p_{\mathcal{C}^{(4)}\ast}r^{(4)}=0,
\end{equation}
where $r^{(4)}$ is the ramification divisor given by
$r^{(4)}=p_{\mathcal{C}^{(4)}\ast}c_1-c_1(\mathcal{C}^{(4)})$. It
is convenient to define the cover flux $\gamma^{(4)}$ by
\begin{equation}
c_1(\mathcal{L})=\lambda\gamma^{(4)}+\frac{1}{2}r^{(4)},
\end{equation}
where $\lambda$ is a rational number used to compensate the
non-integral class $\frac{1}{2}r^{(4)}$ such that
$c_1(\mathcal{L})\in H_4({\bar X},\mathbb{Z})$. The traceless
condition $c_1(p_{\mathcal{C}^{(4)}\ast}\mathcal{L})=0$ is then
equivalent to the condition
$p_{\mathcal{C}^{(4)}\ast}\gamma^{(4)}=0$. Up to multiplication of
a constant, the only choice of $\gamma^{(4)}$ satisfying the
traceless condition is
\begin{equation}
\gamma^{(4)}=(4-p^{\ast}_{\mathcal{C}^{(4)}}p_{\mathcal{C}^{(4)}\ast})([\mathcal{C}^{(4)}]\cdot\sigma).
\end{equation}
Since the first Chern class of a line bundle must be integral, it
follows that $\lambda$ and $\gamma^{(4)}$ have to obey the
following quantization condition
\begin{equation}
\lambda
\gamma^{(4)}+\frac{1}{2}[p_{\mathcal{C}^{(4)}}^{\ast}c_1-c_1(\mathcal{C}^{(4)})]\in
H_{4}(\bar X,\mathbb{Z}).
\end{equation}
With the given cover flux $\gamma^{(4)}$, the net chirality of
matter ${\bf 16}$ is calculated by \cite{Donagi:global,
Other:global}
\begin{equation}
N_{\bf
16}=([\mathcal{C}^{(4)}]\cdot\sigma)\cdot\lambda\gamma^{(4)} =
-\lambda\eta\cdot(\eta-4c_1).\label{chirality semi field 16}
\end{equation}
On the other hand, the homological class of matter $\bf 10$ curve
is given by Eq.~(\ref{D curve}). It turns out that the net
chirality of matter ${\bf 10}$ is computed as \cite{Other:global}
\begin{eqnarray}
N_{\bf 10}=[D]\cdot\gamma^{(4)}=0.\label{chirality semi field 10}
\end{eqnarray}
One can find that the computations of net chirality agree with those
from heterotic spectral cover. Unlike the representation $\bf 10$ in $SU(5)$ case, the $\bf 10$ in $SO(10)$ is a real representation. Therefore,
it is impossible to engineer a chiral spectrum of $\bf 10$'s by using a generic $SU(4)$ spectral cover. From Eq.~(\ref{chirality semi field
16}) and Eq.~(\ref{chirality semi field 10}), we obtain an
$SO(10)$ model with $-\lambda\eta\cdot(\eta-4c_1)$ copies of
matter on the ${\bf 16}$ curve and nothing on the $\bf 10$ curve.
The flux does not have many degrees of freedom to tune and the
candidate of $\bf 10$ Higgs is absent. Therefore, we shall consider factorizations of the $SU(4)$
cover $\mathcal{C}^{(4)}$ to enrich the configuration along the
line of the $SU(5)$ cover studied in \cite{Caltech:global02,
Caltech:global03, Blumenhagen:global01, Blumenhagen:global02}.
Before studying the cove factorizations, we shall construct an
$SU(4)$ spectral divisor motivated from heterotic/F-theory
duality\cite{Marsano:2010ix} in section 3.2.

\subsection{$SU(4)$ Spectral Cover Divisor}

Recall that $Z_4$ is an elliptically fibered Calabi-Yau fourfold
$\pi:Z_4\ra B_3$ with a section $\sigma: B_3\ra Z_4$. In general,
$Z_4$ can be described by the Weierstrass model\footnote{Recall
that $Z_4$ can be embedded as a hypersurface of
$W\mathbb{P}^3_{2,3,1}$ -fibration over $B_3$. It follows from the
Calabi-Yau condition $c_1(Z_4)=0$ that $x$, $y$, and $u$ are
sections of $\mathcal{O}_{B_3}(2\sigma)\otimes K_{B_3}^{-2}$,
$\mathcal{O}_{B_3}(3\sigma)\otimes K_{B_3}^{-3}$, and
$\mathcal{O}_{B_3}(\sigma)$, respectively.}
\begin{equation}
y^2=x^3+fxu^4+gu^6,\label{Weiestrassmodel}
\end{equation}
where $f$ and $g$ are sections of $K_{B_3}^{-4}$ and
$K_{B_3}^{-6}$, respectively. We now consider the case that $Z_4$
exhibits a $D_5$ singularity along a divisor $B_2$ inside $B_3$.
We define $z$ to be a section of the normal bundle $N_{B_2/B_3}$
of $B_2$ in $B_3$. Locally we can expand $f$ and $g$ in the
Weierstrass model Eq.~(\ref{Weiestrassmodel}) in terms of $z$.
With suitable choice of variables, we obtain
\begin{equation}
y^2=x^3+u(zu)[b_0(zu)^4+b_2(zu)^2x+b_3(zu)y+b_4x^2]+\mathcal{O}(z;u),\label{Weiestrassmodel
with su4}
\end{equation}
where $\mathcal{O}(z;u)$ stands for the higher order terms of $z$
for each fixed order in $u$. Following the proposal in
\cite{Marsano:2010ix}, we define the $SU(4)$ spectral divisor as
\begin{equation}
\mathcal{D}^{(4)}:\;b_0(zu)^4+b_2(zu)^2x+b_3(zu)y+b_4x^2=0\label{spectral
divisor su4}
\end{equation}
with a projection map $p_{\mathcal{D}^{(4)}}: \mathcal{D}^{(4)}\ra
B_2$. Note that local behavior of $\mathcal{D}^{(4)}$ is the same
as the union of the exceptional lines described by
Eq.~(\ref{except lines SU4}) and that the homological classes of
$x$, $y$, $u$, $z$, and $b_m$ in Eq.~(\ref{spectral divisor su4})
are
\begin{equation}
[x]=2[\sigma+\pi^{\ast}c_1(B_3)],\;\;[y]=3[\sigma+\pi^{\ast}c_1(B_3)],\;\;[u]=\sigma,\;\;[z]=\pi^{\ast}B_2
\end{equation}
\begin{equation}
[b_m]=(6-m)\pi^{\ast}c_1(B_3)-(5-m)\pi^{\ast}B_2,\;\;\; m=0,2,3,4.
\end{equation}
The homological class of the divisor $\mathcal{D}^{(4)}$ is then
given by
\begin{equation}
[\mathcal{D}^{(4)}]=4\sigma+\pi^{\ast}[6c_1(B_3)-B_2].\label{class
su4 cover}
\end{equation}
In this case the dual matter ${\bf 16}$ surface
$\widehat{\Sigma}_{\bf 16}$ is determined by the locus of
$\{(zu)=0\}\cap\{b_4=0\}$ with homological class
\begin{equation}
[\widehat{\Sigma}_{\bf
16}]=(\sigma+\pi^{\ast}B_2)\cdot\pi^{\ast}[2c_1(B_3)-B_2].
\end{equation}
On the other hand, the dual matter ${\bf 10}$ surface sits inside
the locus of the intersection
$\mathcal{D}^{(4)}\cap\tau\mathcal{D}^{(4)}$, where $\tau$ is a
$\mathbb{Z}_2$ involution acting on the cover by $y\ra -y$ while
keeping $x$, $u$, and $z$ invariant. More precisely, the
intersection $\mathcal{D}^{(4)}\cap\tau\mathcal{D}^{(4)}$ is given
by
\begin{equation}
\left\{\begin{array}{l} b_3(zu)y=0\\
b_0(zu)^4+b_2(zu)^2x+b_4x^2=0.
\end{array}\label{matter 10 cover}   \right.
\end{equation}
We can compute the homological class $[\widehat{\Sigma}_{\bf 10}]$
of dual matter ${\bf 10}$ surface as
\begin{eqnarray}
[\widehat{\Sigma}_{\bf
10}]&=&[\mathcal{D}^{(4)}]\cdot[\mathcal{D}^{(4)}]-[zu]\cdot[b_4]-[y]\cdot[b_4x^2]-2[x]\cdot[z]\nonumber\\&=&\sigma\cdot\pi^{\ast}[12c_1(B_3)-8B_2]
+\pi^{\ast}[3c_1(B_3)-2B_2]\cdot\pi^{\ast}[6c_1(B_3)-B_2].
\end{eqnarray}
To obtain chiral spectrum, we turn on a spectral line bundle
$\mathcal{N}$ over $\mathcal{D}^{(4)}$. The corresponding Higgs
bundle is given by $E=p_{\mathcal{D}^{(4)}\ast}\mathcal{N}$. For
$SU(n)$ bundles, it is required that $c_1(E)=0$. It follows that
\begin{equation}
c_1(p_{\mathcal{D}^{(4)}\ast}\mathcal{N})=p_{\mathcal{D}^{(4)}\ast}c_1(\mathcal{N})-\frac{1}{2}p_{\mathcal{D}^{(4)}\ast}\widehat{r}^{(4)}=0,
\end{equation}
where $\widehat{r}^{(4)}$ is the ramification divisor given by
$\widehat{r}^{(4)}=p_{\mathcal{D}^{(4)}\ast}c_1(B_3)-c_1(\mathcal{D}^{(4)})$.
It is convenient to define the flux $\widehat{\gamma}^{(4)}$ by
\begin{equation}
c_1(\mathcal{N})=\lambda\widehat{\gamma}^{(4)}+\frac{1}{2}\widehat{r}^{(4)},
\end{equation}
where $\lambda$ is a rational number used to compensate the
non-integral class $\frac{1}{2}\widehat{r}^{(4)}$ such that
$c_1(\mathcal{N})\in H_2(\mathcal{D}^{(4)},\mathbb{Z})$. The
traceless condition $c_1(p_{\mathcal{D}^{(4)}\ast}\mathcal{N})=0$
is then equivalent to the condition
$p_{\mathcal{D}^{(4)}\ast}\widehat{\gamma}^{(4)}=0$. Up to
multiplication of a constant, the only choice of
$\widehat{\gamma}^{(4)}$ satisfying the traceless condition is
\begin{equation}
\widehat{\gamma}^{(4)}=(4-p^{\ast}_{\mathcal{D}^{(4)}}p_{\mathcal{D}^{(4)}\ast})([\mathcal{D}^{(4)}]\cdot\sigma).
\end{equation}
Since the first Chern class of a line bundle must be integral, it
follows that $\lambda$ and $\widehat{\gamma}^{(4)}$ have to obey
the following quantization condition
\begin{equation}
\lambda
\widehat{\gamma}^{(4)}+\frac{1}{2}[p_{\mathcal{D}^{(4)}}^{\ast}c_1(B_3)-c_1(\mathcal{D}^{(4)})]\in
H_{2}(\mathcal{D}^{(4)},\mathbb{Z}).
\end{equation}
In the case of $SU(4)$ spectral divisor, the traceless flux
$\widehat{\gamma}^{(4)}$ is given by
\begin{equation}
\widehat{\gamma}^{(4)}=(4-p_{\mathcal{D}^{(4)}}^{\ast}p_{{\mathcal{D}^{(4)}}\ast})([\mathcal{D}^{(4)}]\cdot\sigma)=[\mathcal{D}^{(4)}]\cdot\{4\sigma-\pi^{\ast}[2c_1(B_3)-B_2]\}.
\label{Twist on d4}
\end{equation}
It follows from Eq.~(\ref{Twist on d4}) and the definition
$\widehat{\gamma}^{(4)}=[\mathcal{D}^{(4)}]\cdot
\mathcal{G}^{(4)}$ that
$\mathcal{G}^{(4)}=4\sigma-\pi^{\ast}[2c_1(B_3)-B_2]$. With the
given cover flux $\widehat{\gamma}^{(4)}$, the net chirality of
matter ${\bf 16}$ and ${\bf 10}$ are respectively given by
\begin{eqnarray}
N_{\bf 16}&=&[\widehat{\Sigma}_{\bf
16}]\cdot\mathcal{G}^{(4)}\cdot\pi^{\ast}B_2\nonumber\\&=&-(6c_1-t)\cdot_{B_2}(2c_1-t),
\end{eqnarray}
and
\begin{eqnarray}
N_{\bf 10}&=&[\widehat{\Sigma}_{\bf
10}]\cdot\mathcal{G}^{(4)}\cdot\pi^{\ast}B_2\nonumber\\&=&0,
\end{eqnarray}
where the fact that $B_2|_{B_2}=-t$ and $c_1(B_3)|_{B_2}=c_1-t$
has been used. We found agreement between net chirality from
semi-local spectral cover and from spectral divisor construction.

\section{Chirality}

In this section we consider flipped $SU(5)$ GUTs in F-theory. As
mentioned in section 1, the construction contains two steps. The
first step is to break $E_8$ down to $SO(10)$ by using $SU(4)$
spectral covers. The second step is to turn on $U(1)_X$ fluxes to
break $SO(10)$ down to $SU(5)\times U(1)_X$. In what follows we
shall focus on the first step, namely breaking $E_8$ down to
$SO(10)$ by using a semi-local $SU(4)$ spectral cover and its
global completion, $SU(4)$ spectral divisors. We also analyze the
chiral spectra induced by the fluxes. For the analysis of $U(1)_X$
fluxes and numerical models, we refer readers to\cite{Chen:2010tp}
for the details. We first briefly review $(3,1)$ and $(2,2)$
factorizations of the semi-local $SU(4)$ spectral cover and induced
chirality. Then we construct the factorized $SU(4)$ spectral
divisor for each factorization and calculate the chirality induced
by the fluxes.

\subsection{Semi-local $SU(4)$ Spectral Cover}

\subsubsection{Constraints}
Before computing the chiral spectra, we take a moment to analyze
the constraints for the cover fluxes. Let us consider the case of
the cover factorization
$\mathcal{C}^{(n)}\ra\mathcal{C}^{(l)}\times\mathcal{C}^{(m)}$. To
obtain well-defined cover fluxes and maintain supersymmetry, we
impose the following constraints \cite{Caltech:global03}:
\begin{eqnarray}
&&c_1(p_{{\mathcal{C}^{(l)}}\ast}\mathcal{L}^{(l)}) +
c_1(p_{{\mathcal{C}^{(m)}}\ast}\mathcal{L}^{(m)})
=0,\label{cover constraint 01}\\
&&c_1(\mathcal{L}^{(k)})\in
H_2(\mathcal{C}^{(k)},\mathbb{Z}),\;\;k=l,m,\label{cover constraint
02}\\
&&[c_1(p_{{\mathcal{C}^{(l)}}\ast}\mathcal{L}^{(l)}) -
c_1(p_{{\mathcal{C}^{(m)}}\ast}\mathcal{L}^{(m)})]\cdot_{B_2}
[\omega]=0,\label{cover constraint 03}
\end{eqnarray}
where $p_{\mathcal{C}^{(k)}}$ denotes the projection map
$p_{\mathcal{C}^{(k)}}:\mathcal{C}^{(k)}\rightarrow B_2$,
$\mathcal{L}^{(k)}$ is a line bundle over $\mathcal{C}^{(k)}$ and
$[\omega]$ is an ample divisor dual to a K\"ahler form of $B_2$.
The first constraint Eq.~(\ref{cover constraint 01}) is the
traceless condition for the induced Higgs bundle\footnote{We may
think of Eq.~(\ref{cover constraint 02}) as the traceless
condition of an $SU(4)$ bundle $V_4$ over $B_2$ split into
$V_3\oplus L$ with $V_3=p_{a\ast}\mathcal{L}^{(a)}$ and
$L=p_{b\ast}\mathcal{L}^{(b)}$. Therefore, the traceless condition
of $V_4$ can be expressed by $c_1(V_4)
=c_1(p_{a\ast}\mathcal{L}^{(a)}) +
c_1(p_{b\ast}\mathcal{L}^{(b)})=0$.}. The second constraint
Eq.~(\ref{cover constraint 02}) requires that the first Chern
class of a well-defined line bundle $\mathcal{L}^{(k)}$ must be
integral. The third constraint states that the 2-cycle
$[c_1(p_{{\mathcal{C}^{(l)}}\ast}\mathcal{L}^{(l)})-c_1(p_{{\mathcal{C}^{(m)}}\ast}\mathcal{L}^{(m)})]$
in $B_2$ is supersymmetic. Note that Eq.~(\ref{cover constraint
01}) can be expressed as
\begin{equation}
p_{{\mathcal{C}^{(l)}}\ast}c_1(\mathcal{L}^{(l)})-\frac{1}{2}p_{{\mathcal{C}^{(l)}}\ast}r^{(l)}
+
p_{{\mathcal{C}^{(m)}}\ast}c_1(\mathcal{L}^{(m)})-\frac{1}{2}p_{{\mathcal{C}^{(m)}}\ast}r^{(m)}=0,
\end{equation}
where $r^{(l)}$ and $r^{(m)}$ are the ramification divisors for
the maps $p_{\mathcal{C}^{(l)}}$ and $p_{\mathcal{C}^{(m)}}$,
respectively. Recall that the ramification divisor $r^{(k)}$ is
defined by
\begin{equation}
r^{(k)}=p_{\mathcal{C}^{(k)}}^{\ast} c_1
-c_1(\mathcal{C}^{(k)}),\;\;k=l,m.\label{r divisor}
\end{equation}
It is convenient to define cover fluxes $\gamma^{(k)}$ as
\begin{equation}
c_1(\mathcal{L}^{(k)})=\gamma^{(k)}+\frac{1}{2}
r^{(k)},\;\;k=l,m.\label{cover flux def}
\end{equation}
With Eq.~(\ref{cover flux def}), the traceless condition
Eq.~(\ref{cover constraint 01}) can be expressed as
$p_{{\mathcal{C}^{(l)}}\ast}\gamma^{(l)}+p_{{\mathcal{C}^{(m)}}\ast}\gamma^{(m)}=0$.
By using Eq.~(\ref{r divisor}) and Eq.~(\ref{cover flux def}), we
can recast the quantization condition Eq.~(\ref{cover constraint
02}) by
$\gamma^{(k)}+\frac{1}{2}[p^{\ast}_{\mathcal{C}^{(k)}}c_1-c_1(\mathcal{C}^{(k)})]\in
H_2(\mathcal{C}^{(k)},\mathbb{Z}),\;\;k=l,m$. It follows from
Eq.~(\ref{cover constraint 01}) that the condition Eq.~(\ref{cover
constraint 03}) can be reduced to
$p_{{\mathcal{C}^{(k)}}\ast}\gamma^{(k)}\cdot_{B_2}[\omega]=0$. We
summarize the constraints for the cover fluxes $\gamma^{(k)}$ as
follows:
\begin{eqnarray}
&&
p_{{\mathcal{C}^{(l)}}\ast}\gamma^{(l)}+p_{{\mathcal{C}^{(m)}}\ast}\gamma^{(m)}=0,\label{simplified
cover constraint 01}
\\
&&\gamma^{(k)}+\frac{1}{2}[p^{\ast}_{\mathcal{C}^{(k)}}c_1-c_1(\mathcal{C}^{(k)})]\in
H_2(\mathcal{C}^{(k)},\mathbb{Z}),\;\;k=l,m,\label{simplified cover
constraint 02}
\\
&&
p_{{\mathcal{C}^{(k)}}\ast}\gamma^{(k)}\cdot_{B_2}[\omega]=0,\;\;k=l,m.\label{simplified
cover constraint 03}
\end{eqnarray}
In the next section, we shall calculate the homological classes of
matter curves for $(3,1)$ and $(2,2)$ factorizations. We also
compute the chirality induced by the restriction of the fluxes to
each matter curve.

\subsubsection{$(3,1)$ Factorization}

We consider the $(3,1)$ factorization, $\mathcal{C}^{(4)}
\rightarrow \mathcal{C}^{(a)}\times\mathcal{C}^{(b)} $
corresponding to the factorization of Eq.~(\ref{homo SU(4) cover})
as follows:
\begin{equation}
\mathcal{C}^{(a)}\times\mathcal{C}^{(b)}:\;\;(a_0 U^3 + a_1U^2W
+a_2 UW^2 + a_3 W^3)(d_0 U +d_1W)=0.
\end{equation}
By comparing with Eq.~(\ref{homo SU(4) cover}), we can obtain the
following decomposition:
\begin{equation}
b_0 = a_0 d_0,~~ b_1 = a_1 d_0 + a_0 d_1=0,~~ b_2 = a_2 d_0 + a_1
d_1,~~ b_3 = a_3 d_0 + a_2 d_1,~~ b_4 = a_3 d_1.\label{coef 3,1
facorization}
\end{equation}
We denote the classes $[d_1]$ by $\pi^{\ast}\xi_1$ and then write
\begin{equation}
[d_0]=\pi^{\ast}(c_1+\xi_1),~~ ~~ [a_k]=\pi^{\ast}[\eta
-(k+1)c_1-\xi_1],\;\;\;k=0,1,2,3.
\end{equation}
To solve the traceless condition $b_1=0$, we use ansatz $a_0=\alpha d_0$ and $a_1=-\alpha d_1$ where $[\alpha]=\pi^{\ast}(\eta-2c_1-2\xi_1)$.
It is easy to see that the homological classes of
$\mathcal{C}^{(a)}$ and $\mathcal{C}^{(b)}$ in $\bar X$ are
\begin{equation}
[\mathcal{C}^{(a)}]= 3\sigma+\pi^{\ast}(\eta
-c_1-\xi_1),\;\;\;[\mathcal{C}^{(b)}]=\sigma+\pi^{\ast}(c_1+\xi_1).\label{31
Classes}
\end{equation}

To obtain the ${\bf 10}$ curves, we follow the method
proposed in \cite{Donagi:2004ia, Caltech:global02,
Caltech:global03, Blumenhagen:global01} to calculate the
intersection $\mathcal{C}^{(4)}\cap \tau \mathcal{C}^{(4)}$, where
$\tau$ is the $\mathbb{Z}_2$ involution $\tau: W\rightarrow -W$
acting on the spectral cover. Since the calculation is
straightforward, we omit the detailed calculation here and only
summarize the results in Table
\ref{3-1 10curves}\footnote{To simplify notations, we denote
$\mathcal{C}^{(k)}\cap\tau\mathcal{C}^{(l)}$ by
$\mathcal{C}^{(k)(l)}$ and notice that
$[\mathcal{C}^{(k)(l)}]=[\mathcal{C}^{(l)(k)}]$.}\footnote{To avoid a singularity of non-Kodaira type, we impose the condition $\xi_1\cdot_{B_2} (c_1+\xi_1)=0$.
Therefore, $[\Sigma_{{\bf 10}^{(b)(b)}}]=\pi^{\ast}
\xi_1\cdot\pi^{\ast} (c_1+\xi_1)$=0.}.

\begin{table}[h]
\begin{center}
\renewcommand{\arraystretch}{.75}
\begin{tabular}{|c|c|c|c|} \hline
& $[\mathcal{C}^{(b)(b)}]$ & $2[\mathcal{C}^{(a)(b)}]$ & $[\mathcal{C}^{(a)(a)}]$ \\
\hline

{\bf 16} & $\sigma\cdot\pi^{\ast}\xi_1$ & - &
$\sigma\cdot\pi^{\ast} (\eta-4c_1-\xi_1)$ \\ \hline

\multirow{2}{*}{\bf 10} & \multirow{2}{*}{-} &
$2[\sigma+\pi^{\ast}(c_1+\xi_1)]$ &
$[2\sigma+\pi^{\ast}(\eta-2c_1 -\xi_1)]$\\
& & $\cdot~\pi^{\ast}(\eta-3c_1-\xi_1)+2\sigma\cdot
\pi^{\ast}\xi_1$ & $\cdot~\pi^{\ast}(\eta
-3c_1-\xi_1)+2(\sigma+\pi^{\ast}c_1)\cdot \pi^{\ast} \xi_1$ \\
\hline

\multirow{2}{*}{$\infty$} & \multirow{2}{*}{$\sigma_{\infty}
\cdot\pi^{\ast} (c_1+\xi_1)$} & \multirow{2}{*}{4$\sigma_{\infty}
\cdot \pi^{\ast} (c_1+\xi_1)$} & $\sigma_{\infty} \cdot
\pi^{\ast}(\eta-c_1 -\xi_1)$ \\
& & & $+2\sigma_{\infty} \cdot \pi^{\ast}(\eta -2c_1-2\xi_1)$
\\ \hline
\end{tabular}
\caption{Matter curves for the factorization $\mathcal{C}^{(4)}=
\mathcal{C}^{(a)}\times\mathcal{C}^{(b)}$.} \label{3-1 10curves}
\end{center}
\end{table}

It follows from Table \ref{3-1 10curves} that the homological
classes 
of $\bf 16$ curves are
\begin{eqnarray}
&&[\Sigma_{{\bf 16}^{(a)}}]=\sigma\cdot\pi^{\ast} (\eta-4c_1-\xi_1)~~~~ \\
&&[\Sigma_{{\bf 16}^{(b)}}]=\sigma\cdot\pi^{\ast}\xi_1
\end{eqnarray}
and that the homological classes of $[\Sigma_{{\bf 10}^{(a)(a)}}]$ and $[\Sigma_{{\bf
10}^{(a)(b)}}]$ are\footnote{It follows from Eqs.~(\ref{aacurve}) and (\ref{abcurve}) that $[\Sigma_{{\bf 10}^{(a)(a)}}]$ and $[\Sigma_{{\bf 10}^{(a)(b)}}]$
correspond to the same matter curve in $B_2$ with homological class $\eta-3c_1$. In other words, $\Sigma_{{\bf 10}^{(a)(a)}}$ and $\Sigma_{{\bf 10}^{(a)(b)}}$ both are
lifts of the same curve in $B_2$. The $\bf 10$ matter curve inside the cover $\mathcal{C}^{(4)}$ is actually 4-sheeted cover of the corresponding matter curve in $B_2$.
A nice description of the cover structure for the $\bf 10$ curve can be found in \cite{Other:global}.}
\begin{eqnarray}
&&[\Sigma_{{\bf 10}^{(a)(a)}}]=[2\sigma+\pi^{\ast}(\eta-2c_1
-\xi_1)]\cdot~\pi^{\ast}(\eta
-3c_1-\xi_1)+2(\sigma+\pi^{\ast}c_1)\cdot \pi^{\ast} \xi_1~~~~ \label{aacurve}\\
&&[\Sigma_{{\bf
10}^{(a)(b)}}]=[\sigma+\pi^{\ast}(c_1+\xi_1)]\cdot~\pi^{\ast}(\eta-3c_1-\xi_1)+\sigma\cdot
\pi^{\ast}\xi_1.\label{abcurve}
\end{eqnarray}

For the $(3,1)$ factorization, the ramification divisors for the
spectral covers $\mathcal{C}^{(a)}$ and $\mathcal{C}^{(b)}$ are
given by
\begin{eqnarray}
&&r^{(a)}=[\mathcal{C}^{(a)}]\cdot[\sigma+\pi^{\ast}(\eta-2c_1-\xi_1)] \nonumber\\
&&r^{(b)}=[\mathcal{C}^{(b)}]\cdot (-\sigma+\pi^{\ast}\xi_1),
\end{eqnarray}
respectively. We define traceless fluxes $\gamma^{(a)}_0$ and
$\gamma^{(b)}_0$ by
\begin{eqnarray}
&&\gamma^{(a)}_0=(3-p_{\mathcal{C}^{(a)}}^{\ast}
p_{\mathcal{C}^{(a)}\ast}) \gamma^{(a)} =[\mathcal{C}^{(a)}]\cdot
[ 3
\sigma-\pi^{\ast}(\eta-4c_1-\xi_1)]\nonumber\\
&&\gamma^{(b)}_0=(1-p_{\mathcal{C}^{(b)}}^{\ast}
p_{\mathcal{C}^{(b)}\ast})\gamma^{(b)} =[\mathcal{C}^{(b)}]\cdot
\left( \sigma-\pi^{\ast}\xi_1\right),\label{Flux_01_3,1}
\end{eqnarray}
where $\gamma^{(a)}$ and $\gamma^{(b)}$ are non-traceless fluxes
and defined by
\begin{equation}
\gamma^{(a)}=[\mathcal{C}^{(a)}]\cdot \sigma,~~
\gamma^{(b)}=[\mathcal{C}^{(b)}]\cdot \sigma.
\end{equation}
Then we can  calculate the restriction of fluxes $\gamma^{(a)}_0$
and $\gamma^{(b)}_0$ to each matter curve. We omit the calculation
here and only summarize the results in Table \ref{Chirality 3-1
gamma}.
\begin{table}[h]
\begin{center}
\renewcommand{\arraystretch}{.75}
\begin{tabular}{|c|c|c|c|} \hline
& $\gamma^{(b)}_0$ & $\gamma^{(a)}_0$ \\\hline

${\bf 16}^{(b)}$ & $-\xi_1\cdot_{B_2}(c_1+\xi_1)$ & 0 \\ \hline

${\bf 16}^{(a)}$ & 0 &
$-(\eta-c_1-\xi_1)\cdot_{B_2}(\eta-4c_1-\xi_1)$
\\\hline

${\bf 10}^{(a)(b)}$ & 0 &
$-(\eta-3c_1-3\xi_1)\cdot_{B_2}(\eta-4c_1-\xi_1)$ \\\hline

${\bf 10}^{(a)(a)}$ & 0 &
$(\eta-3c_1-3\xi_1)\cdot_{B_2}(\eta-4c_1-\xi_1)$
\\\hline
\end{tabular}
\caption{Chirality induced by the fluxes $\gamma^{(a)}_0$ and
$\gamma^{(b)}_0$.} \label{Chirality 3-1 gamma}
\end{center}
\end{table}
We also can define additional fluxes $\delta^{(a)}$ and
$\delta^{(b)}$ by
\begin{eqnarray}
&&\delta^{(a)}=(1-p_{\mathcal{C}^{(b)}}^{\ast}
p_{\mathcal{C}^{(a)}\ast}) \gamma^{(a)} =[\mathcal{C}^{(a)}]\cdot
\sigma-[\mathcal{C}^{(b)}]\cdot \pi^{\ast}(\eta-4c_1-\xi_1)
\nonumber
\\&&\delta^{(b)}=(3-p_{\mathcal{C}^{(a)}}^{\ast} p_{\mathcal{C}^{(b)}\ast}) \gamma^{(b)}
=[\mathcal{C}^{(b)}]\cdot 3\sigma-
[\mathcal{C}^{(a)}]\cdot\pi^{\ast}\xi_1. \label{Flux_02_3,1}
\end{eqnarray}
Another flux we can include is \cite{Caltech:global03}
\begin{equation}
{\rho}^{(3,1)}=(3p_{\mathcal{C}^{(b)}}^{\ast}-p_{\mathcal{C}^{(a)}}^{\ast})\rho,\label{Flux_03_3,1}
\end{equation}
where $\rho\in H_2(B_2,\mathbb{R})$.  We summarize the restriction
of fluxes $\delta^{(a)}$, $\delta^{(b)}$ and ${\rho}^{(3,1)}$ to
each matter curve in Table \ref{Chirality 3-1 delta rho}.

\begin{table}[h]
\begin{center}
\renewcommand{\arraystretch}{.75}
\begin{tabular}{|c|c|c|c|} \hline
& $\delta^{(b)}$ & $\delta^{(a)}$ & ${\rho}^{(3,1)}$ \\\hline

${\bf 16}^{(b)}$ & $-3c_1\cdot_{B_2}\xi_1$ &
$-\xi_1\cdot_{B_2}(\eta-4c_1-\xi_1)$ & $3\rho\cdot_{B_2} \xi_1$ \\
\hline

${\bf 16}^{(a)}$ & $-\xi_1\cdot_{B_2} (\eta-4c_1-\xi_1)$ &
$-c_1\cdot_{B_2}(\eta-4c_1-\xi_1)$ & $-\rho\cdot_{B_2}
(\eta-4c_1-\xi_1)$
\\\hline

${\bf 10}^{(a)(b)}$ & $\xi_1\cdot_{B_2}(2\eta-9c_1-3\xi_1)$ &
$-(\eta-3c_1-\xi_1)\cdot_{B_2}(\eta-4c_1-\xi_1)$&
$2\rho\cdot_{B_2}(\eta-3c_1)$
\\\hline

${\bf 10}^{(a)(a)}$ & $-2\xi_1\cdot_{B_2} (\eta-3c_1)$ &
$(\eta-3c_1-\xi_1)\cdot_{B_2}(\eta-4c_1-\xi_1)$ &
$-2\rho\cdot_{B_2} (\eta-3c_1)$
\\\hline
\end{tabular}
\caption{Chirality induced by the fluxes $\delta^{(a)}$,
$\delta^{(b)}$, and ${\rho}^{(3,1)}$.} \label{Chirality 3-1 delta
rho}
\end{center}
\end{table}
With Eqs.~(\ref{Flux_01_3,1}), (\ref{Flux_02_3,1}), and
(\ref{Flux_03_3,1}), we define the universal cover flux $\Gamma$
to be \cite{Caltech:global03}
\begin{equation}
\Gamma=k_a \gamma^{(a)}_0+k_b\gamma^{(b)}_0 +m_a \delta^{(a)} +m_b
\delta^{(b)} + {\rho}^{(3,1)}\equiv\Gamma^{(a)}+\Gamma^{(b)},
\end{equation}
where $\Gamma^{(a)}$ and $\Gamma^{(b)}$ are defined by
\begin{eqnarray}
&&\Gamma^{(a)}=[\mathcal{C}^{(a)}]\cdot \left[(3k_a+m_a)\sigma
-\pi^{\ast}(k_a (\eta-4c_1-\xi_1) +m_b\xi_1 + \rho)\right],
\\&&\Gamma^{(b)}=[\mathcal{C}^{(b)}]\cdot \left[(k_b+3m_b)\sigma
-\pi^{\ast}(k_b\xi_1+m_a(\eta-4c_1-\xi_1)-3\rho) \right].
\end{eqnarray}
Note that
\begin{eqnarray}
&&p_{\mathcal{C}^{(a)}\ast}\Gamma^{(a)}= -3m_b\xi_1+m_a(\eta-4c_1-\xi_1)-3\rho,\\
&&p_{\mathcal{C}^{(b)}\ast} \Gamma^{(b)}= 3m_b\xi_1
-m_a(\eta-4c_1-\xi_1)+3\rho.
\end{eqnarray}
Clearly, $\Gamma^{(a)}$ and $\Gamma^{(b)}$ obey the traceless
condition
$p_{\mathcal{C}^{(a)}\ast}\Gamma^{(a)}+p_{\mathcal{C}^{(b)}\ast}\Gamma^{(b)}=0$.
Besides, the quantization condition in this case becomes
\begin{equation}
(3k_a+m_a+\frac{1}{2})\sigma -\pi^{\ast}[k_a(\eta-4c_1-\xi_1)+
m_b\xi_1 + \rho-\frac{1}{2}(\eta-2c_1-\xi_1)] \in H_4({\bar
X},\mathbb{Z}),
\end{equation}
\begin{equation}
(k_b+3m_b-\frac{1}{2})\sigma -\pi^{\ast}[k_b\xi_1
+m_a(\eta-4c_1-\xi_1)-3\rho -\frac{1}{2}\xi_1] \in H_4({\bar
X},\mathbb{Z}).
\end{equation}
The supersymmetry condition is given by
\begin{equation}
[3m_b\xi_1-m_a(\eta-4c_1-\xi_1)+3\rho]\cdot_{B_2}[\omega]=0.\label{BPS
3,1}
\end{equation}

\subsubsection{$(2,2)$ Factorization}

In the case of the (2,2) factorization, the cover is split as
$\mathcal{C}^{(4)} \rightarrow \mathcal{C}^{(d_1)}\times
\mathcal{C}^{(d_2)}$. We then can factorize Eq.~(\ref{homo SU(4)
cover}) into the following form:
\begin{equation}
\mathcal{C}^{(d_1)}\times \mathcal{C}^{(d_2)}:\;\;\;(e_0U^2 +
e_1UW +e_2W^2)(f_0U^2 + f_1UW + f_2W^2) =0
\end{equation}
By comparing the coefficients with Eq.~(\ref{homo SU(4) cover}),
we obtain
\begin{equation}
b_0=e_0f_0,~~ b_1=e_0f_1+e_1f_0=0,~~ b_2=e_0f_2+e_1f_1+e_2f_0,~~
b_3 = e_1f_2+ e_2f_1,~~ b_4=e_2f_2.
\end{equation}
By denoting the homological class of $f_2$ by $\pi^{\ast}\xi_2$,
the classes of other sections can be written as
\begin{equation}
[f_1]= \pi^{\ast}(c_1+\xi_2), ~~ [f_0]= \pi^{\ast}(2c_1+\xi_2),
~~[e_m]=\pi^{\ast}[ \eta-(m+2)c_1 - \xi_2],\;\;\;m=0,1,2.
\end{equation}
To solve the traceless condition $b_1=0$, we impose the condition $e_0=\beta f_0$ and $e_1=-\beta f_1$ where $[\beta]=\pi^{\ast}(\eta-4c_1-2\xi_2)$.
In this case, the homological classes of $\mathcal{C}^{(d_1)}$ and
$\mathcal{C}^{(d_2)}$ are given by
\begin{equation}
[\mathcal{C}^{(d_1)}]= 2\sigma+\pi^{\ast}(\eta
-2c_1-\xi_2),\;\;\;[\mathcal{C}^{(d_2)}]=2\sigma+\pi^{\ast}(2c_1+\xi_2).
\end{equation}
To find the ${\bf 10}$ curves, we again
follow the method proposed in \cite{Donagi:2004ia,
Caltech:global02, Caltech:global03, Blumenhagen:global01} to
calculate the intersection $\mathcal{C}^{(4)}\cap \tau
\mathcal{C}^{(4)}$. We omit the detailed calculation here and only
summarize the
results 
in Table \ref{2-2 10curves}.
\begin{table}[h]
\begin{center}
\renewcommand{\arraystretch}{.75}
\begin{tabular}{|c|c|c|c|} \hline
& $[\mathcal{C}^{(d_2)(d_2)}]$ & $2[\mathcal{C}^{(d_1)(d_2)}]$ & $[\mathcal{C}^{(d_1)(d_1)}]$ \\
\hline

{\bf 16} & $\sigma\cdot\pi^{\ast}\xi_2$ & - &
$\sigma\cdot\pi^{\ast} (\eta-4c_1-\xi_2)$ \\ \hline

\multirow{2}{*}{\bf 10} & $[2\sigma +\pi^{\ast}( 2c_1+\xi_2)]$ &
$2[2\sigma+\pi^{\ast}(2c_1+\xi_2)]$ &
$\pi^{\ast}(\eta-3c_1-\xi_2)\cdot \pi^{\ast}(\eta-4c_1 -\xi_2)$\\
&$\cdot~\pi^{\ast} (c_1+\xi_2)$ &
$\cdot~\pi^{\ast}(\eta-4c_1-\xi_2)$ & $+2(\sigma+\pi^{\ast}c_1
)\cdot \pi^{\ast}(c_1+\xi_2)$ \\ \hline

\multirow{2}{*}{$\infty$} & \multirow{2}{*}{$\sigma_{\infty}
\cdot\pi^{\ast} (2c_1+\xi_2)$} & \multirow{2}{*}{4$\sigma_{\infty}
\cdot \pi^{\ast} (2c_1+\xi_2)$} & $\sigma_{\infty} \cdot
\pi^{\ast}(\eta-2c_1 -\xi_2)$ \\
& & & $+2\sigma_{\infty} \cdot \pi^{\ast}(\eta -4c_1-2\xi_2)$
\\ \hline
\end{tabular}
\caption{Matter curves for the factorization
$\mathcal{C}^{(4)}=\mathcal{C}^{(d_1)}\times
\mathcal{C}^{(d_2)}$.} \label{2-2 10curves}
\end{center}
\end{table}

It follows from Table \ref{2-2 10curves} that the homological
classes of the factorized $\bf 16$ curves are
\begin{eqnarray}
&&[\Sigma_{{\bf 16}^{(d_1)}}]=\sigma\cdot\pi^{\ast} (\eta-4c_1-\xi_2),~~~~~~~~ \\
&&[\Sigma_{{\bf 16}^{(d_2)}}]=\sigma\cdot\pi^{\ast}\xi_2,
\end{eqnarray}
and that the homological classes of the factorized $\bf 10$ curves
are\footnote{It follows from Eqs.~(\ref{d1d1curve})-(\ref{d2d2curve}) that $[\Sigma_{{\bf 10}^{(d_1)(d_1)}}]$ and $[\Sigma_{{\bf 10}^{(d_2)(d_2)}}]$ correspond to
the same curve with class $c_1+\xi_2$ in $B_2$, and $[\Sigma_{{\bf 10}^{(d_1)(d_2)}}]|_{\sigma}=2(\eta-4c_1-\xi_2)$ in $B_2$.}
\begin{eqnarray}
&&[\Sigma_{{\bf 10}^{(d_1)(d_1)}}]=2(\sigma+\pi^{\ast}c_1 )\cdot
\pi^{\ast}(c_1+\xi_2)+\pi^{\ast}(\eta-3c_1-\xi_2)\cdot
\pi^{\ast}(\eta-4c_1 -\xi_2),~~~~~~~~ \label{d1d1curve}\\
&&[\Sigma_{{\bf
10}^{(d_1)(d_2)}}]=[2\sigma+\pi^{\ast}(2c_1+\xi_2)]\cdot~\pi^{\ast}(\eta-4c_1-\xi_2),\label{d1d2curve}
\\
&&[\Sigma_{{\bf
10}^{(d_2)(d_2)}}]=[2\sigma+\pi^{\ast}(2c_1+\xi_2)]\cdot~\pi^{\ast}(c_1+\xi_2).\label{d2d2curve}
\end{eqnarray}
In the $(2,2)$ factorization, the ramification divisors
$r^{(d_1)}$ and $r^{(d_2)}$ for the covers $\mathcal{C}^{(d_1)}$
and $\mathcal{C}^{(d_2)}$ are given by
\begin{eqnarray}
&&r^{(d_1)}= [\mathcal{C}^{(d_1)}]\cdot\pi^{\ast}(\eta-3c_1-\xi_2), \nonumber\\
&&r^{(d_2)}= [\mathcal{C}^{(d_2)}]\cdot\pi^{\ast}(c_1+\xi_2),
\end{eqnarray}
respectively. We then define traceless cover fluxes
$\gamma^{(d_1)}_0$ and $\gamma^{(d_2)}_0$ by
\begin{eqnarray}
&&\gamma^{(d_1)}_0=(2-p_{\mathcal{C}^{(d_1)}}^{\ast}
p_{\mathcal{C}^{(d_1)}\ast})\gamma^{(d_1)}
=[\mathcal{C}^{(d_1)}]\cdot \left[ 2
\sigma-\pi^{\ast}(\eta-4c_1-\xi_2)\right],\nonumber \\
&&\gamma^{(d_2)}_0=(2-p_{\mathcal{C}^{(d_2)}}^{\ast}
p_{\mathcal{C}^{(d_2)}\ast})\gamma^{(d_2)}
=[\mathcal{C}^{(d_2)}]\cdot \left( 2
\sigma-\pi^{\ast}\xi_2\right),
\end{eqnarray}
where $\gamma^{(d_1)}$ and $\gamma^{(d_21)}$ are non-traceless
fluxes and defined by
\begin{equation}
\gamma^{(d_1)}=[\mathcal{C}^{(d_1)}]\cdot \sigma,~~
\gamma^{(d_2)}=[\mathcal{C}^{(d_2)}]\cdot \sigma.
\end{equation}
We summarize the restriction of the fluxes to each factorized
curve in Table \ref{Chirality 2-2 gamma}.
\begin{table}[h]
\begin{center}
\renewcommand{\arraystretch}{.75}
\begin{tabular}{|c|c|c|c|} \hline
& $\gamma^{(d_2)}_0$ & $\gamma^{(d_1)}_0$ \\\hline

${\bf 16}^{(d_2)}$ & $-\xi_2\cdot_{B_2}(2c_1+\xi_2)$ & 0 \\ \hline

${\bf 16}^{(d_1)}$ & 0 &
$-(\eta-2c_1-\xi_2)\cdot_{B_2}(\eta-4c_1-\xi_2)$ \\\hline

${\bf 10}^{(d_2)(d_2)}$ & 0 & 0 \\ \hline

${\bf 10}^{(d_1)(d_2)}$ & 0 &
$-2(\eta-4c_1-2\xi_2)\cdot_{B_2}(\eta-4c_1-\xi_2)$
\\\hline

${\bf 10}^{(d_1)(d_1)}$ & 0 &
$2(\eta-4c_1-2\xi_2)\cdot_{B_2}(\eta-4c_1-\xi_2)$
\\\hline
\end{tabular}
\caption{Chirality induced by the fluxes $\gamma^{(d_1)}_0$ and
$\gamma^{(d_2)}_0$.} \label{Chirality 2-2 gamma}
\end{center}
\end{table}
We also can define two fluxes
\begin{eqnarray}
&&\delta^{(d_1)}=(2-p_{\mathcal{C}^{(d_2)}}^{\ast}
p_{\mathcal{C}^{(d_1)}\ast}) \gamma^{(d_1)}
=[\mathcal{C}^{(d_1)}]\cdot
2\sigma- [\mathcal{C}^{(d_2)}]\cdot\pi^{\ast}(\eta-4c_1-\xi_2), \nonumber \\
&&\delta^{(d_2)}=(2-p_{\mathcal{C}^{(d_1)}}^{\ast}
p_{\mathcal{C}^{(d_2)}\ast}) \gamma^{(d_2)}
=[\mathcal{C}^{(d_2)}]\cdot 2\sigma-[\mathcal{C}^{(d_1)}]\cdot
\pi^{\ast}\xi_2.
\end{eqnarray}
Another flux we can include is \cite{Caltech:global03}
\begin{equation}
{\rho}^{(2,2)}=(p_{\mathcal{C}^{(d_2)}}^{\ast}-p_{\mathcal{C}^{(d_1)}}^{\ast})\rho,
\end{equation}
where $\rho\in H_2(B_2,\mathbb{R})$. We summarize the restriction
of the fluxes $\delta^{(d_1)}$, $\delta^{(d_2)}$, and
${\rho}^{(2,2)}$ to each factorized curve in Table \ref{Chirality
2-2 delta rho}.
\begin{table}[h]
\begin{center}
\renewcommand{\arraystretch}{.75}
\begin{tabular}{|c|c|c|c|} \hline
& $\delta^{(d_2)}$ & $\delta^{(d_1)}$ & ${\rho}^{(2,2)}$
\\\hline

${{\bf 16}^{(d_2)}}$ & $-2c_1\cdot_{B_2}\xi_2$ &
$-\xi_2\cdot_{B_2} (\eta-4c_1-\xi_2)$ & $\rho\cdot_{B_2} \xi_2$ \\
\hline

${{\bf 16}^{(d_1)}}$ & $-\xi_2\cdot_{B_2} (\eta-4c_1-\xi_2)$ &
$-2c_1\cdot_{B_2}(\eta-4c_1-\xi_2)$ & $-\rho\cdot_{B_2}
(\eta-4c_1-\xi_2)$
\\\hline

${{\bf 10}^{(d_2)(d_2)}}$ & $2\xi_2\cdot_{B_2} (c_1+\xi_2)$ &
$-2(c_1+\xi_2)\cdot_{B_2}(\eta-4c_1-\xi_2)$ & $2\rho\cdot_{B_2}(c_1+\xi_2)$\\
\hline

${{\bf 10}^{(d_1)(d_2)}}$ & 0 &
$-2(\eta-4c_1-2\xi_2)\cdot_{B_2}(\eta-4c_1-\xi_2)$&0
\\\hline

${{\bf 10}^{(d_1)(d_1)}}$ & $-2\xi_2\cdot_{B_2}(c_1+\xi_2)$ &
$2(\eta-3c_1-\xi_2)\cdot_{B_2}(\eta-4c_1-\xi_2)$ &
$-2\rho\cdot_{B_2} (c_1+\xi_2)$
\\\hline
\end{tabular}
\caption{Chirality induced by the fluxes $\delta^{(d_1)}$,
$\delta^{(d_2)}$, and ${\rho}^{(2,2)}$.} \label{Chirality 2-2
delta rho}
\end{center}
\end{table}

Again we conclude the universal cover flux to be
\begin{equation}
\Gamma=k_{d_1} \gamma^{(d_1)}_0+k_{d_2}\gamma^{(d_2)}_0 +m_{d_1}
\delta^{(d_1)} +m_{d_2} \delta^{(d_2)} +
{\rho}^{(2,2)}=\Gamma^{(d_1)}+\Gamma^{(d_2)},
\end{equation}
where
\begin{eqnarray}
&&\Gamma^{(d_1)}=[\mathcal{C}^{(d_1)}]\cdot
\left\{2(k_{d_1}+m_{d_1})\sigma -\pi^{\ast}[k_{d_1}
(\eta-4c_1-\xi_2)+m_{d_2}\xi_2+ \rho]\right\},\nonumber \\
&&\Gamma^{(d_2)}=[\mathcal{C}^{(d_2)}]\cdot
\left\{2(k_{d_2}+m_{d_2})\sigma
-\pi^{\ast}[k_{d_2}\xi_2+m_{d_1}(\eta-4c_1-\xi_2)-\rho] \right\} .
\end{eqnarray}
Note that
\begin{eqnarray}
&&p_{\mathcal{C}^{(d_1)}\ast} \Gamma^{(d_1)}
=-2m_{d_2}\xi_2+2m_{d_1}(\eta-4c_1-\xi_2)-2\rho,\\
&&p_{\mathcal{C}^{(d_2)}\ast} \Gamma^{(d_2)} =
2m_{d_2}\xi_2-2m_{d_1}(\eta-4c_1-\xi_2)+2\rho.
\end{eqnarray}
It is easy to see that $\Gamma^{(d_1)}$ and $\Gamma^{(d_2)}$
satisfy the traceless condition $p_{\mathcal{C}^{(d_1)}\ast}
\Gamma^{(d_1)}+p_{\mathcal{C}^{(d_2)}\ast} \Gamma^{(d_2)}=0$. In
addition, the quantization condition in this case becomes
\begin{equation}
2(k_{d_1}+m_{d_1})\sigma -\pi^{\ast}[k_{d_1}(\eta-4c_1-\xi_2)+
m_{d_2}\xi_2 + \rho-\frac{1}{2}(\eta-3c_1-\xi_2)] \in H_4({\bar
X},\mathbb{Z}),
\end{equation}
\begin{equation}
2(k_{d_2}+m_{d_2})\sigma
-\pi^{\ast}[k_{d_2}\xi_2+m_{d_1}(\eta-4c_1-\xi_2)-\rho
-\frac{1}{2}(c_1+\xi_2)]\in H_4({\bar X},\mathbb{Z}).
\end{equation}
The supersymmetry condition is then given by
\begin{equation}
[2m_{d_2}\xi_2-2m_{d_1}(\eta-4c_1-\xi_2)+2\rho]\cdot_{B_2}[\omega]=0.\label{BPS
2,2}
\end{equation}

\subsection{Global $SU(4)$ Spectral Divisor}

\subsubsection{Constraints}

Similar to the analysis in the last section, we analyze the
constraints for the fluxes of the spectral divisors. It was argued
in\cite{Marsano:2010ix} that these constraints could be consistent
with that for the semi-local cover fluxes. Let us consider the
case of the cover factorization
$\mathcal{D}^{(n)}\ra\mathcal{D}^{(l)}\times\mathcal{D}^{(m)}$. To
obtain well-defined cover fluxes and maintain supersymmetry, we
impose the following constraints \cite{Marsano:2010ix}:
\begin{eqnarray}
&&c_1(p_{{\mathcal{D}^{(l)}}\ast}\mathcal{N}^{(l)}) +
c_1(p_{{\mathcal{D}^{(m)}}\ast}\mathcal{N}^{(m)})
=0,\label{cover constraint 001}\\
&&c_1(\mathcal{N}^{(k)})\in
H_2(\mathcal{D}^{(k)},\mathbb{Z}),\;\;k=l,m,\label{cover constraint
002}
\end{eqnarray}
where $p_{\mathcal{D}^{(k)}}$ denotes the projection map
$p_{\mathcal{D}^{(k)}}:\mathcal{D}^{(k)}\rightarrow B_3$,
$\mathcal{N}^{(k)}$ is a line bundle over $\mathcal{C}^{(k)}$. The
first constraint, Eq.~(\ref{cover constraint 001}) is the
traceless condition for the induced Higgs bundle. The second
constraint, Eq.~(\ref{cover constraint 002}) requires that the
first Chern class of a well-defined line bundle
$\mathcal{N}^{(k)}$ must be integral. Note that Eq.~(\ref{cover
constraint 001}) can be expressed as
\begin{equation}
p_{{\mathcal{D}^{(l)}}\ast}c_1(\mathcal{N}^{(l)})-\frac{1}{2}p_{{\mathcal{D}^{(l)}}\ast}\widehat{r}^{(l)}
+
p_{{\mathcal{D}^{(m)}}\ast}c_1(\mathcal{N}^{(m)})-\frac{1}{2}p_{{\mathcal{D}^{(m)}}\ast}\widehat{r}^{(m)}=0,
\end{equation}
where $\widehat{r}^{(l)}$ and $\widehat{r}^{(m)}$ are the
ramification divisors for the maps $p_{\mathcal{D}^{(l)}}$ and
$p_{\mathcal{D}^{(m)}}$, respectively. Recall that the
ramification divisor $\widehat{r}^{(k)}$ is defined by
\begin{equation}
\widehat{r}^{(k)}=p_{\mathcal{D}^{(k)}}^{\ast} c_1(B_3)
-c_1(\mathcal{D}^{(k)}),\;\;k=l,m.\label{r divisor01}
\end{equation}
It is convenient to define fluxes $\widehat{\gamma}^{(k)}$ as
\begin{equation}
c_1(\mathcal{N}^{(k)})=\widehat{\gamma}^{(k)}+\frac{1}{2}
\widehat{r}^{(k)},\;\;k=l,m.\label{cover flux def02}
\end{equation}
With Eq.~(\ref{cover flux def02}), the traceless condition
Eq.~(\ref{cover constraint 001}) can be expressed as
$p_{{\mathcal{D}^{(l)}}\ast}\widehat{\gamma}^{(l)}+p_{{\mathcal{D}^{(m)}}\ast}\widehat{\gamma}^{(m)}=0$.
By using Eq.~(\ref{r divisor01}) and Eq.~(\ref{cover flux def02}),
we can recast the quantization condition Eq.~(\ref{cover
constraint 002}) by
$\widehat{\gamma}^{(k)}+\frac{1}{2}[p^{\ast}_{\mathcal{D}^{(k)}}c_1(B_3)-c_1(\mathcal{D}^{(k)})]\in
H_2(\mathcal{D}^{(k)},\mathbb{Z}),\;\;k=l,m$. We summarize the
constraints for the fluxes $\widehat{\gamma}^{(k)}$ as follows:
\begin{eqnarray}
&&
p_{{\mathcal{D}^{(l)}}\ast}\widehat{\gamma}^{(l)}+p_{{\mathcal{D}^{(m)}}\ast}\widehat{\gamma}^{(m)}=0\label{simplified
cover constraint 001}
\\
&&\widehat{\gamma}^{(k)}+\frac{1}{2}[p^{\ast}_{\mathcal{D}^{(k)}}c_1(B_3)-c_1(\mathcal{D}^{(k)})]\in
H_2(\mathcal{D}^{(k)},\mathbb{Z}),\;\;k=l,m.\label{simplified cover
constraint 002}
\end{eqnarray}
In the next section, we shall calculate the homological classes of
the dual matter surfaces for $(3,1)$ and $(2,2)$ factorizations.
We also compute the chirality induced by the restriction of the
fluxes to each dual matter surface.

\subsubsection{$(3,1)$ Factorization}

It will be convenient to define $x=\zeta^2$ and $y=\zeta^3$ where
$\zeta$ is a section of $\mathcal{O}_{B_3}(\sigma)\otimes
K_{B_3}^{-1}$. Then the $SU(4)$ spectral divisor defined by Eq.
(\ref{spectral divisor su4}) can be written as
\begin{equation}
\mathcal{D}^{(4)}:\;\;
b_0(zu)^4+b_2(zu)^2\zeta^2+b_3(zu)\zeta^3+b_4\zeta^4=0.\label{C_Zeta}
\end{equation}
We now consider the $(3,1)$ factorization $\mathcal{D}^{(4)}
\rightarrow \mathcal{D}^{(a)}\times\mathcal{D}^{(b)} $
corresponding to the factorization of Eq. (\ref{C_Zeta})
\begin{equation}
\mathcal{D}^{(a)}\times\mathcal{D}^{(b)}:\;\;[\widetilde{a}_0
(zu)^3 + \widetilde{a}_1(zu)^2\zeta +\widetilde{a}_2 (zu)\zeta^2 +
\widetilde{a}_3 \zeta^3][\widetilde{d}_0 (zu)
+\widetilde{d}_1\zeta]=0,\label{3,1 global factorization}
\end{equation}
with projection maps $p_{\mathcal{D}^{(a)}}:\mathcal{D}^{(a)}\ra
B_3$ and $p_{\mathcal{D}^{(b)}}:\mathcal{D}^{(b)}\ra B_3$. By
comparing with Eq. (\ref{C_Zeta}), we can obtain the following
relations:
\begin{equation}
b_0 = \widetilde{a}_0 \widetilde{d}_0,~~ b_1 = \widetilde{a}_1
\widetilde{d}_0 + \widetilde{a}_0 \widetilde{d}_1=0,~~ b_2 =
\widetilde{a}_2 \widetilde{d}_0 + \widetilde{a}_1\widetilde{
d}_1,~~ b_3 = \widetilde{a}_3 \widetilde{d}_0 + \widetilde{a}_2
\widetilde{d}_1,~~ b_4 = \widetilde{a}_3
\widetilde{d}_1.\label{coef 3,1 global facorization}
\end{equation}
We denote the homological class of $[\widetilde{d}_1]$ by
$\pi^{\ast}\widehat{\xi}_1$ and then write
\begin{equation}
[\widetilde{d}_0]=\pi^{\ast}[c_1(B_3)-B_2+\widehat{\xi}_1],~~ ~~
[\widetilde{a}_k]=\pi^{\ast
}[(5-m)c_1(B_3)-(4-m)B_2-\widehat{\xi}_1],\;\;m=0,1,2,3.
\end{equation}
It is easy to see that the homological classes of
$\mathcal{D}^{(a)}$ and $\mathcal{D}^{(b)}$ are given by
\begin{equation}
[\mathcal{D}^{(a)}]=
3\sigma+\pi^{\ast}[5c_1(B_3)-B_2-\widehat{\xi}_1],\;\;\;\;[\mathcal{D}^{(b)}]=\sigma+\pi^{\ast}[c_1(B_3)+\widehat{\xi}_1].\label{31
global Classes}
\end{equation}
Note that the unfactorized dual matter ${\bf 16}$ surface sits
inside the locus of $\{(zu)=0\}\cap\{b_4=0\}$. Due to the
factorization in Eq.~(\ref{3,1 global factorization}), the
factorized dual matter ${\bf 16}$ surfaces sit inside the loci
$\{(zu)=0\}\cap\{\widetilde{a}_3=0\}$ and
$\{(zu)=0\}\cap\{\widetilde{d}_1=0\}$. The homological class of
dual matter surfaces $\widehat{\Sigma}_{{\bf 16}^{(a)}}$ and
$\widehat{\Sigma}_{{\bf 16}^{(b)}}$ are given by
\begin{equation}
[\widehat{\Sigma}_{{\bf
16}^{(a)}}]=(\sigma+\pi^{\ast}B_2)\cdot\pi^{\ast}[2c_1(B_3)-B_2-\widehat{\xi}_1],\;\;\;[\widehat{\Sigma}_{{\bf
16}^{(b)}}]=(\sigma+\pi^{\ast}B_2)\cdot\pi^{\ast}\widehat{\xi}_1.
\end{equation}
To obtain dual matter surface $\widehat{\Sigma}_{\bf 10}$'s, we
calculate the intersection $\mathcal{D}^{(4)}\cap\tau
\mathcal{D}^{(4)}$, where $\tau$ is a $\mathbb{Z}_2$ involution
$\zeta\ra -\zeta$ acting on $\mathcal{D}^{(4)}$
\cite{Donagi:2004ia, Caltech:global02, Caltech:global03,
Blumenhagen:global01}. Under $(3,1)$ factorization
$\mathcal{D}^{(4)}\ra \mathcal{D}^{(a)}\times \mathcal{D}^{(b)}$,
the intersection $\mathcal{D}^{(4)}\cap\tau \mathcal{D}^{(4)}$ can
be decomposed into several components $\mathcal{D}^{(a)}\cap\tau
\mathcal{D}^{(a)}$, $\mathcal{D}^{(a)}\cap\tau \mathcal{D}^{(b)}$,
and $\mathcal{D}^{(b)}\cap\tau \mathcal{D}^{(b)}$. We first
consider the case of $\mathcal{D}^{(a)}\cap\tau
\mathcal{D}^{(a)}$. This intersection is determined by
\begin{equation}
\left\{\begin{array}{l} (zu)[\widetilde{a}_0(zu)^2+\widetilde{a}_2\zeta^2]=0\\
\zeta[\widetilde{a}_1(zu)^2+\widetilde{a}_3\zeta^2]=0.
\end{array}\label{matter 10 cover}   \right.
\end{equation}
To solve the constraint $b_1=\widetilde{a}_1\widetilde{d}_0 +
\widetilde{a}_0 \widetilde{d}_1=0$, we use ansatz
$\widetilde{a}_0=\widetilde{\alpha} \widetilde{d}_0$ and
$\widetilde{a}_1=-\widetilde{\alpha }\widetilde{d}_1$ where the
homological class of $\widetilde{\alpha}$ is
$[\widetilde{\alpha}]=\pi^{\ast}[4c_1(B_3)-3B_2-2\widehat{\xi}_1]$.
By using the ansatz, we obtain
\begin{equation}
\left\{\begin{array}{l} (zu)[\widetilde{\alpha} \widetilde{d}_0(zu)^2+\widetilde{a}_2\zeta^2]=0\\
\zeta[-\widetilde{\alpha
}\widetilde{d}_1(zu)^2+\widetilde{a}_3\zeta^2]=0.
\end{array}\label{Fact DaDa component}   \right.
\end{equation}
It follows from Eq. (\ref{Fact DaDa component}) that the
homological class of dual matter surface $\widehat{\Sigma}_{{\bf
10}^{(a)(a)}}$ is given by
\begin{eqnarray}
[\widehat{\Sigma}_{{\bf 10}^{(a)(a)}}]&=&[\mathcal{D}^{(a)}]\cdot
[\mathcal{D}^{(a)}]-[\zeta]\cdot [\widetilde{a}_0]-[zu]\cdot
[\widetilde{a}_3]-9[\zeta]\cdot [zu]-2[\zeta]\cdot
[\widetilde{\alpha}]
\nonumber\\&=&\{2\sigma+\pi^{\ast}[4c_1(B_3)-B_2-\widehat{\xi}_1]\}\cdot\pi^{\ast}[3c_1(B_3)-2B_2-\widehat{\xi}_1]
\nonumber\\&+&2[\sigma+\pi^{\ast}c_1(B_3)]\cdot\pi^{\ast}\widehat{\xi}_1.
\end{eqnarray}
Next we calculate the intersection $\mathcal{D}^{(a)}\cap\tau
\mathcal{D}^{(b)}$ which is given by
\begin{equation}
\left\{\begin{array}{l} \widetilde{a}_0(zu)^3+\widetilde{a}_1(zu)^2\zeta+\widetilde{a}_2(zu)\zeta^2+\widetilde{a}_3\zeta^3=0\\
\widetilde{d}_0(zu)-\widetilde{d}_1\zeta=0.
\end{array}\label{Fact DaDb component 00}   \right.
\end{equation}
By using the ansatz, we can rewrite Eq. (\ref{Fact DaDb component
00}) as
\begin{equation}
\left\{\begin{array}{l} \zeta^2[\widetilde{a}_2(zu)+\widetilde{a}_3\zeta]=0\\
\widetilde{d}_0(zu)-\widetilde{d}_1\zeta=0.
\end{array}\label{Fact DaDb component}   \right.
\end{equation}
It follows from Eq. (\ref{Fact DaDb component}) that the
homological class of dual matter surface $\widehat{\Sigma}_{{\bf
10}^{(a)(b)}}$ is
\begin{eqnarray}
[\widehat{\Sigma}_{{\bf 10}^{(a)(b)}}]&=&[\mathcal{D}^{(a)}]\cdot
[\mathcal{D}^{(b)}]-2[\zeta]\cdot [\widetilde{d}_0]-3[\zeta]\cdot
[zu]\nonumber\\&=&\{\sigma+\pi^{\ast}[c_1(B_3)+\widehat{\xi}_1]\}\cdot\pi^{\ast}[3c_1(B_3)-2B_2-\widehat{\xi}_1]\nonumber\\&+&
(\sigma+\pi^{\ast}B_2)\cdot\pi^{\ast}\widehat{\xi}_1.
\end{eqnarray}
Let us turn to the case of $\mathcal{D}^{(b)}\cap\tau
\mathcal{D}^{(b)}$ which is determined by
\begin{equation}
\left\{\begin{array}{l} \widetilde{d}_0(zu)=0\\
\widetilde{d}_1\zeta=0.
\end{array}\label{matter 10 cover}   \right.
\end{equation}
Then the homological class of dual matter surface
$\widehat{\Sigma}_{{\bf 10}^{(b)(b)}}$ is given by
\begin{eqnarray}
[\widehat{\Sigma}_{{\bf 10}^{(b)(b)}}]&=&[\mathcal{D}^{(b)}]\cdot
[\mathcal{D}^{(b)}]-[\zeta]\cdot [\widetilde{d}_0]-[zu]\cdot
[\widetilde{d}_1]-[\zeta]\cdot[zu]\nonumber\\&=&\pi^{\ast}[c_1(B_3)-B_2+\widehat{\xi}_1]\cdot\pi^{\ast}\widehat{\xi}_1.
\end{eqnarray}
We summarize the homological classes of dual matter $\bf 16$ and
${\bf 10}$ surfaces in Table \ref{Dual Surface 3-1}\footnote{In the case of ${\bf 10}^{(b)(b)}$,
we impose the condition $\pi^{\ast}[c_1(B_3)-B_2+\widehat{\xi}_1]\cdot\pi^{\ast}\widehat{\xi}_1=0$ to avoid the appearance of a singularity.}.
\begin{table}[h]
\begin{center}
\renewcommand{\arraystretch}{.75}
\begin{tabular}{|c|c|c|c|} \hline
Field & Homological Class  \\\hline

${\bf 16}^{(b)}$ & $(\sigma+\pi^{\ast}B_2)\cdot\pi^{\ast}\widehat{\xi}_1$ \\
\hline

${\bf 16}^{(a)}$ &
$(\sigma+\pi^{\ast}B_2)\cdot\pi^{\ast}[2c_1(B_3)-B_2-\widehat{\xi}_1]$
\\\hline

${\bf 10}^{(b)(b)}$ &
-
\\\hline

${\bf 10}^{(a)(b)}$ &
$\{\sigma+\pi^{\ast}[c_1(B_3)+\widehat{\xi}_1]\}\cdot\pi^{\ast}[3c_1(B_3)-2B_2-\widehat{\xi}_1]$\\

& $+(\sigma+\pi^{\ast}B_2)\cdot\pi^{\ast}\widehat{\xi}_1$\\\hline

 ${\bf 10}^{(a)(a)}$ & $\{2\sigma+\pi^{\ast}[4c_1(B_3)-B_2-\widehat{\xi}_1]\}\cdot\pi^{\ast}[3c_1(B_3)-2B_2-\widehat{\xi}_1]$
\\
&
$+2[\sigma+\pi^{\ast}c_1(B_3)]\cdot\pi^{\ast}\widehat{\xi}_1$\\\hline

\end{tabular}
\caption{Dual matter surfaces for the factorization
$\mathcal{D}^{(4)}= \mathcal{D}^{(a)}\times\mathcal{D}^{(b)}$.}
\label{Dual Surface 3-1}
\end{center}
\end{table}

In $(3,1)$ factorization, the ramification divisors for
$\mathcal{D}^{(a)}$ and $\mathcal{D}^{(b)}$ are given by
\begin{eqnarray}
&&\widehat{r}^{(a)}=[\mathcal{D}^{(a)}]\cdot\{\sigma+\pi^{\ast}[4c_1(B_3)-2B_2-\widehat{\xi}_1]\},\nonumber \\
&&\widehat{r}^{(b)}=[\mathcal{D}^{(b)}]\cdot
[-\sigma-\pi^{\ast}(B_2-\widehat{\xi}_1)],
\end{eqnarray}
respectively. We define traceless fluxes
$\widehat{\gamma}^{(a)}_0$ and $\widehat{\gamma}^{(b)}_0$ by
\begin{eqnarray}
&&\widehat{\gamma}^{(a)}_0=(3-p_{\mathcal{D}^{(a)}}^{\ast}
p_{\mathcal{D}^{(a)}\ast}) \widehat{\gamma}^{(a)}
=[\mathcal{D}^{(a)}]\cdot \{ 3
\sigma-\pi^{\ast}[2c_1(B_3)-B_2-\widehat{\xi}_1]\},\nonumber\\
&&\widehat{\gamma}^{(b)}_0=(1-p_{\mathcal{D}^{(b)}}^{\ast}
p_{\mathcal{D}^{(b)}\ast})\widehat{\gamma}^{(b)}
=[\mathcal{D}^{(b)}]\cdot (
\sigma-\pi^{\ast}\widehat{\xi}_1),\label{Flux_1_3,1}
\end{eqnarray}
where $\widehat{\gamma}^{(a)}$ and $\widehat{\gamma}^{(b)}$ are
non-traceless fluxes and defined by
\begin{equation}
\widehat{\gamma}^{(a)}=[\mathcal{D}^{(a)}]\cdot \sigma,~~
\widehat{\gamma}^{(b)}=[\mathcal{D}^{(b)}]\cdot \sigma.
\end{equation}
Following the formula in section 3.2, the net chirality of matter
in the representation $\mathbf{r}$ induced by the flux
$\mathcal{G}$ is
\begin{equation}
N_{\mathbf{r}}=[\widehat{\Sigma}_{\mathbf{r}}]\cdot\mathcal{G}\cdot\pi^{\ast}B_2,\label{Net
Chiralty formula}
\end{equation}
where $[\widehat{\Sigma}_{\mathbf{r}}]$ is the homological class
of dual surface for matter in the representation ${\mathbf{r}}$.
By using Eq. (\ref{Net Chiralty formula}) and
$\widehat{\xi}_1|_{B_2}=\xi_1$, we can calculate the restriction
of fluxes $\widehat{\gamma}^{(a)}_0$ and
$\widehat{\gamma}^{(b)}_0$ to each dual matter surface. We omit
the calculation here and only summarize the results in Table
\ref{Chirality global 3-1 gamma}.
\begin{table}[h]
\begin{center}
\renewcommand{\arraystretch}{.75}
\begin{tabular}{|c|c|c|c|} \hline
& $\widehat{\gamma}^{(b)}_0$ & $\widehat{\gamma}^{(a)}_0$ \\\hline

${\bf 16}^{(b)}$ & $-\xi_1\cdot_{B_2}(c_1+\xi_1)$ & 0 \\ \hline

${\bf 16}^{(a)}$ & 0 & $-(5c_1-t-\xi_1)\cdot_{B_2}(2c_1-t-\xi_1)$
\\\hline

${\bf 10}^{(a)(b)}$ & 0 &
$-(3c_1-t-3\xi_1)\cdot_{B_2}(2c_1-t-\xi_1)$ \\\hline

${\bf 10}^{(a)(a)}$ & 0 &
$(3c_1-t-3\xi_1)\cdot_{B_2}(2c_1-t-\xi_1)$\\\hline
\end{tabular}
\caption{Chirality induce by the fluxes $\widehat{\gamma}^{(a)}_0$
and $\widehat{\gamma}^{(b)}_0$.} \label{Chirality global 3-1
gamma}
\end{center}
\end{table}

We also can define additional fluxes $\widehat{\delta}^{(a)}$ and
$\widehat{\delta}^{(b)}$ by
\begin{eqnarray}
&&\widehat{\delta}^{(a)}=(1-p_{\mathcal{D}^{(b)}}^{\ast}
p_{\mathcal{D}^{(a)}\ast}) \widehat{\gamma}^{(a)}
=[\mathcal{D}^{(a)}]\cdot \sigma-[\mathcal{D}^{(b)}]\cdot
\pi^{\ast}[2c_1(B_3)-B_2-\widehat{\xi}_1], \nonumber
\\&&\widehat{\delta}^{(b)}=(3-p_{\mathcal{D}^{(a)}}^{\ast} p_{\mathcal{D}^{(b)}\ast})\widehat{ \gamma}^{(b)}
=[\mathcal{D}^{(b)}]\cdot 3\sigma-
[\mathcal{D}^{(a)}]\cdot\pi^{\ast}\widehat{\xi}_1.
\label{Flux_2_3,1}
\end{eqnarray}
Another flux we can include is \cite{Caltech:global03}
\begin{equation}
\widehat{{\rho}}^{(3,1)}=(3p_{\mathcal{D}^{(b)}}^{\ast}-p_{\mathcal{D}^{(a)}}^{\ast})\widehat{\rho},\label{Flux_3_3,1}
\end{equation}
where $\widehat{\rho}\in H_2(B_3,\mathbb{R})$ with
$\widehat{\rho}|_{B_2}=\rho$. We summarize the restriction of
fluxes $\widehat{\delta}^{(a)}$, $\widehat{\delta}^{(b)}$ and
$\widehat{{\rho}}^{(3,1)}$ to each matter curve in Table
\ref{Chirality global 3-1 delta rho}.
\begin{table}[h]
\begin{center}
\renewcommand{\arraystretch}{.75}
\begin{tabular}{|c|c|c|c|} \hline
 & $\widehat{\delta}^{(b)}$ &
$\widehat{\delta}^{(a)}$ & $\widehat{{\rho}}^{(3,1)}$ \\\hline

${\bf 16}^{(b)}$ & $-3c_1\cdot_{B_2}\xi_1$ &
$-\xi_1\cdot_{B_2}(2c_1-t-\xi_1)$ & $3\rho\cdot_{B_2} \xi_1$ \\
\hline

${\bf 16}^{(a)}$ & $-\xi_1\cdot_{B_2} (2c_1-t-\xi_1)$ &
$-c_1\cdot_{B_2}(2c_1-t-\xi_1)$ & $-\rho\cdot_{B_2}
(2c_1-t-\xi_1)$
\\\hline

${\bf 10}^{(a)(b)}$ & $\xi_1\cdot_{B_2}(3c_1-2t-3\xi_1)$ &
$-(3c_1-t-\xi_1)\cdot_{B_2}(2c_1-t-\xi_1)$&
$2\rho\cdot_{B_2}(3c_1-t)$
\\\hline

${\bf 10}^{(a)(a)}$ & $-2\xi_1\cdot_{B_2} (3c_1-t)$ &
$(3c_1-t-\xi_1)\cdot_{B_2}(2c_1-t-\xi_1)$ & $-2\rho\cdot_{B_2}
(3c_1-t)$
\\\hline
\end{tabular}
\caption{Chirality induce by the fluxes $\widehat{\delta}^{(a)}$,
$\widehat{\delta}^{(b)}$, and $\widehat{{\rho}}^{(3,1)}$.}
\label{Chirality global 3-1 delta rho}
\end{center}
\end{table}

With Eq. (\ref{Flux_1_3,1}), (\ref{Flux_2_3,1}), and
(\ref{Flux_3_3,1}), we define the universal flux
$\widehat{\Gamma}$ to be \cite{Caltech:global03}
\begin{equation}
\widehat{\Gamma}=\widetilde{k}_a
\widehat{\gamma}^{(a)}_0+\widetilde{k}_b\widehat{\gamma}^{(b)}_0
+\widetilde{m}_a \widehat{\delta}^{(a)} +\widetilde{m}_b
\widehat{\delta}^{(b)} +\widehat{
{\rho}}\equiv\widehat{\Gamma}^{(a)}+\widehat{\Gamma}^{(b)},
\end{equation}
where $\widehat{\Gamma}^{(a)}$ and $\widehat{\Gamma}^{(b)}$ are
defined by
\begin{equation}
\widehat{\Gamma}^{(a)}=[\mathcal{D}^{(a)}]\cdot
\{(3\widetilde{k}_a+\widetilde{m}_a)\sigma+\pi^{\ast}[2\widetilde{k}_a
c_1(B_3)-(4\widetilde{k}_a+\widetilde{m}_a)B_2
+(\widetilde{m}_b-\widetilde{k}_a)\widehat{\xi}_1 +
\widehat{\rho}]\},
\end{equation}
\begin{equation}
\widehat{\Gamma}^{(b)}=[\mathcal{D}^{(b)}]\cdot
\{(\widetilde{k}_b+3\widetilde{m}_b)\sigma
-\pi^{\ast}[2\widetilde{m}_a
c_1(B_3)-(\widetilde{k}_b+4\widetilde{m}_b)B_2
+(\widetilde{k}_b-\widetilde{m}_b)\widehat{\xi}_1 -3
\widehat{\rho}] \}.
\end{equation}
Note that
\begin{eqnarray}
&&p_{\mathcal{D}^{(a)}\ast}\widehat{\Gamma}^{(a)}=2\widetilde{m}_ac_1(B_3)-\widetilde{m}_aB_2-(3\widetilde{m}_b+\widetilde{m}_a)\widehat{\xi}_1 -3\widehat{\rho}, \\
&&p_{\mathcal{D}^{(b)}\ast} \widehat{\Gamma}^{(b)}=
-2\widetilde{m}_ac_1(B_3)+\widetilde{m}_aB_2+(3\widetilde{m}_b+\widetilde{m}_a)\widehat{\xi}_1
+3\widehat{\rho}.
\end{eqnarray}
Clearly, $\widehat{\Gamma}^{(a)}$ and $\widehat{\Gamma}^{(b)}$
obey the traceless condition
$p_{\mathcal{D}^{(a)}\ast}\widehat{\Gamma}^{(a)}+p_{\mathcal{D}^{(b)}\ast}\widehat{\Gamma}^{(b)}=0$.
In this case the quantization conditions are
\begin{equation}
\{(3\widetilde{k}_a+\widetilde{m}_a+\frac{1}{2})\sigma+\pi^{\ast}[(2\widetilde{k}_a-1)
c_1(B_3)-(4\widetilde{k}_a+\widetilde{m}_a-1)B_2
+(\widetilde{m}_b-\widetilde{k}_a+\frac{1}{2})\widehat{\xi}_1 +
\widehat{\rho}]\} \in H_4(Z_4,\mathbb{Z}),
\end{equation}
\begin{equation}
\{(\widetilde{k}_b+3\widetilde{m}_b-\frac{1}{2})\sigma
-\pi^{\ast}[2\widetilde{m}_a
c_1(B_3)-(\widetilde{k}_b+4\widetilde{m}_b-\frac{1}{2})B_2
+(\widetilde{k}_b-\widetilde{m}_b-\frac{1}{2})\widehat{\xi}_1 -3
\widehat{\rho}]\} \in H_4(Z_4,\mathbb{Z}).
\end{equation}

\subsubsection{$(2,2)$ Factorization}

In the (2,2) factorization $\mathcal{D}^{(4)} \rightarrow
\mathcal{D}^{(d_1)}\times \mathcal{D}^{(d_2)}$, the divisor
$\mathcal{D}^{(4)}$ splits into two components
$\mathcal{D}^{(d_1)}$ and $\mathcal{D}^{(d_2)}$. We then factorize
Eq. (\ref{C_Zeta}) into the following form:
\begin{equation}
\mathcal{D}^{(d_1)}\times\mathcal{D}^{(d_2)}:\;\;[\widetilde{e}_0
(zu)^2 + \widetilde{e}_1(zu)\zeta
+\widetilde{e}_2\zeta^2][\widetilde{f}_0 (zu)^2 +
\widetilde{f}_1(zu)\zeta +\widetilde{f}_2\zeta^2]=0.
\end{equation}
with projection maps
$p_{\mathcal{D}^{(d_1)}}:\mathcal{D}^{(d_1)}\ra B_3$ and
$p_{\mathcal{D}^{(d_1)}}:\mathcal{D}^{(d_2)}\ra B_3$. By comparing
the coefficients with Eq. (\ref{C_Zeta}), we obtain the following
relations:
\begin{equation}
b_0=\widetilde{e}_0\widetilde{f}_0,~~
b_1=\widetilde{e}_0\widetilde{f}_1+\widetilde{e}_1\widetilde{f}_0=0,~~
b_2=\widetilde{e}_0\widetilde{f}_2+\widetilde{e}_1\widetilde{f}_1+\widetilde{e}_2\widetilde{f}_0,~~
b_3 = \widetilde{e}_1\widetilde{f}_2+
\widetilde{e}_2\widetilde{f}_1,~~
b_4=\widetilde{e}_2\widetilde{f}_2.\label{coef 2,2 global
facorization}
\end{equation}
By denoting the homological class of $\widetilde{f}_2$ by
$\pi^{\ast}\widehat{\xi}_2$, the homological classes of other
sections can be written as
\begin{equation}
[\widetilde{f}_k]=\pi^{\ast}\{(2-k)[c_1(B_3)-B_2]+\widehat{\xi}_2\},\;\;k=0,1,
\end{equation}
\begin{equation}
[\widetilde{e}_m]=\pi^{\ast}[
(m-3)B_2-(m-4)c_1(B_3)-\widehat{\xi}_2],\;\;m=0,1,2.
\end{equation}
In this case, the homological classes of $\mathcal{D}^{(d_1)}$ and
$\mathcal{D}^{(d_2)}$ are given by
\begin{equation}
[\mathcal{D}^{(d_1)}]=
2\sigma+\pi^{\ast}[4c_1(B_3)-B_2-\widehat{\xi}_2],\;\;\;[\mathcal{D}^{(d_2)}]=2\sigma+\pi^{\ast}[2c_1(B_3)+\widehat{\xi}_2].
\end{equation}
With Eq. (\ref{coef 2,2 global facorization}), the dual matter
${\bf 16}$ surfaces sit inside the loci
$\{(zu)=0\}\cap\{\widetilde{e}_2=0\}$ and
$\{(zu)=0\}\cap\{\widetilde{f}_2=0\}$. The homological classes of
dual matter surfaces $\widehat{\Sigma}_{{\bf 16}^{(d_1)}}$ and
$\widehat{\Sigma}_{{\bf 16}^{(d_2)}}$ are given by
\begin{equation}
[\widehat{\Sigma}_{{\bf
16}^{(d_1)}}]=(\sigma+\pi^{\ast}B_2)\cdot\pi^{\ast}[2c_1(B_3)-B_2-\widehat{{\xi}}_2],\;\;\;[\widehat{\Sigma}_{{\bf
16}^{(d_2)}}]=(\sigma+\pi^{\ast}B_2)\cdot\pi^{\ast}\widehat{\xi}_2,
\end{equation}
respectively. We can obtain the homological classes of dual matter
surfaces $\widehat{\Sigma}_{\bf 10}$'s by calculating the
intersection $\mathcal{D}^{(4)}\cap\tau \mathcal{D}^{(4)}$, where
$\tau$ is a $\mathbb{Z}_2$ involution $\zeta\ra -\zeta$
\cite{Donagi:2004ia, Caltech:global02, Caltech:global03,
Blumenhagen:global01}. Under $(2,2)$ factorization
$\mathcal{D}^{(4)}\ra \mathcal{D}^{(d_1)}\times
\mathcal{D}^{(d_2)}$, $\mathcal{D}^{(4)}\cap\tau
\mathcal{D}^{(4)}$ can be decomposed into several components
$\mathcal{D}^{(d_1)}\cap\tau \mathcal{D}^{(d_1)}$,
$\mathcal{D}^{(d_1)}\cap\tau \mathcal{D}^{(d_2)}$, and
$\mathcal{D}^{(d_2)}\cap\tau \mathcal{D}^{(d_2)}$. For the case of
$\mathcal{D}^{(d_1)}\cap\tau \mathcal{D}^{(d_1)}$, this
intersection is determined by
\begin{equation}
\left\{\begin{array}{l}\widetilde{e}_0(zu)^2+\widetilde{e}_2\zeta^2=0\\
\widetilde{e}_1(zu)\zeta=0.
\end{array}\label{Fac Dd1Dd1_00}   \right.
\end{equation}
To solve the constraint
$b_1=\widetilde{e}_0\widetilde{f}_1+\widetilde{e}_1\widetilde{f}_0=0$,
we use ansatz $\widetilde{e}_0=\widetilde{\beta} \widetilde{f}_0$
and $\widetilde{e}_1=-\widetilde{\beta} \widetilde{f}_1$, where
$[\widetilde{\beta}]=\pi^{\ast}[2c_1(B_3)-B_2-2\widehat{\xi}_2]$.
With the ansatz,  Eq. (\ref{Fac Dd1Dd1_00}) can be written as
\begin{equation}
\left\{\begin{array}{l}\widetilde{\beta} \widetilde{f}_0(zu)^2+\widetilde{e}_2\zeta^2=0\\
\widetilde{\beta} \widetilde{f}_1(zu)\zeta=0.
\end{array}\label{Fac Dd1Dd1}   \right.
\end{equation}
It follows from Eq. (\ref{Fac Dd1Dd1}) that the homological class
of dual matter surface $\widehat{\Sigma}_{{\bf 10}^{(d_1)(d_1)}}$
can be computed as
\begin{eqnarray}
[\widehat{\Sigma}_{{\bf
10}^{(d_1)(d_1)}}]&=&[\mathcal{D}^{(d_1)}]\cdot
[\mathcal{D}^{(d_1)}]-[\zeta]\cdot [\widetilde{e}_0]-[zu]\cdot
[\widetilde{e}_2]-4[\zeta]\cdot [zu]-2[\zeta]\cdot
[\beta]\nonumber\\&=&2[\sigma+\pi^{\ast}c_1(B_3)]\cdot\pi^{\ast}[c_1(B_3)-B_2+\widehat{\xi}_2]
\nonumber\\&+&\pi^{\ast}[3c_1(B_3)-2B_2-\widehat{\xi}_2]\cdot\pi^{\ast}[2c_1(B_3)-B_2-\widehat{\xi}_2].
\end{eqnarray}
Next we calculate the intersection $\mathcal{D}^{(d_1)}\cap\tau
\mathcal{D}^{(d_2)}$ which given by
\begin{equation}
\left\{\begin{array}{l} \widetilde{e}_0(zu)^2+\widetilde{e}_1(zu)\zeta+\widetilde{e}_2\zeta^2=0\\
\widetilde{f}_0(zu)^2-\widetilde{f}_1(zu)\zeta+\widetilde{f}_2\zeta^2=0.
\end{array}\label{Fac Dd1Dd2 00}   \right.
\end{equation}
By using the ansatz, we can recast Eq. (\ref{Fac Dd1Dd2 00}) as
\begin{equation}
\left\{\begin{array}{l} \zeta^2[-\beta\widetilde{f}_2+\widetilde{e}_2]=0\\
\widetilde{f}_0(zu)^2-\widetilde{f}_1(zu)\zeta+\widetilde{f}_2\zeta^2=0.
\end{array}\label{Fac Dd1Dd2 }   \right.
\end{equation}
Then the homological class of dual matter surface
$\widehat{\Sigma}_{{\bf 10}^{(d_1)(d_2)}}$ is given by
\begin{eqnarray}
[\widehat{\Sigma}_{{\bf
10}^{(d_1)(d_2)}}]&=&[\mathcal{D}^{(d_1)}]\cdot
[\mathcal{D}^{(d_2)}]-2[\zeta]\cdot
[\widetilde{f}_0]-4[\zeta]\cdot
[zu]\nonumber\\&=&\{2\sigma+\pi^{\ast}[2c_1(B_3)+\widehat{\xi}_2]\}\cdot\pi^{\ast}[2c_1(B_3)-B_2-\widehat{\xi}_2].
\end{eqnarray}
Let us turn to the case of $\mathcal{D}^{(d_2)}\cap\tau
\mathcal{D}^{(d_2)}$. This intersection is described by
\begin{equation}
\left\{\begin{array}{l}\widetilde{f}_0(zu)^2+\widetilde{f}_2\zeta^2=0\\
\widetilde{f}_1(zu)\zeta=0.
\end{array}\label{Fac Dd2Dd2}   \right.
\end{equation}
It follows from Eq. (\ref{Fac Dd2Dd2}) that the homological class
of dual matter surface $\widehat{\Sigma}_{{\bf 10}^{(d_2)(d_2)}}$
is calculated as
\begin{eqnarray}
[\widehat{\Sigma}_{{\bf
10}^{(d_2)(d_2)}}]&=&[\mathcal{D}^{(d_2)}]\cdot
[\mathcal{D}^{(d_2)}]-[\zeta]\cdot [\widetilde{f}_0]-[zu]\cdot
[\widetilde{f}_2]-4[\zeta]\cdot [zu]
\nonumber\\&=&\{2\sigma+\pi^{\ast}[2c_1(B_3)+\widehat{\xi}_2]\}\cdot\pi^{\ast}[c_1(B_3)-B_2+\widehat{\xi}_2].
\end{eqnarray}
We summarize the homological classes of dual matter $\bf 16$ and
${\bf 10}$ surfaces in Table \ref{Dual matter surfaces 2-2}.
\begin{table}[h]
\begin{center}
\renewcommand{\arraystretch}{.75}
\begin{tabular}{|c|c|c|c|} \hline
Field & Homological Class  \\\hline

${\bf 16}^{(d_2)}$ & $(\sigma+\pi^{\ast}B_2)\cdot\pi^{\ast}\widehat{\xi}_2$ \\
\hline

${\bf 16}^{(d_1)}$ &
$(\sigma+\pi^{\ast}B_2)\cdot\pi^{\ast}[2c_1(B_3)-B_2-\widehat{\xi}_2]$
\\\hline

${\bf 10}^{(d_2)(d_2)}$ &
$\{2\sigma+\pi^{\ast}[2c_1(B_3)+\widehat{\xi}_2]\}\cdot\pi^{\ast}[c_1(B_3)-B_2+\widehat{\xi}_2]$
\\\hline

${\bf 10}^{(d_1)(d_2)}$ &
$\{2\sigma+\pi^{\ast}[2c_1(B_3)+\widehat{\xi}_2]\}\cdot\pi^{\ast}[2c_1(B_3)-B_2-\widehat{\xi}_2]$
\\\hline
${\bf 10}^{(d_1)(d_1)}$ &
$2[\sigma+\pi^{\ast}c_1(B_3)]\cdot\pi^{\ast}[c_1(B_3)-B_2+\widehat{\xi}_2]$
\\
&
$+\pi^{\ast}[3c_1(B_3)-2B_2-\widehat{\xi}_2]\cdot\pi^{\ast}[2c_1(B_3)-B_2-\widehat{\xi}_2]$\\\hline
\end{tabular}
\caption{Dual matter surfaces for the factorization
$\mathcal{D}^{(4)}=
\mathcal{D}^{(d_1)}\times\mathcal{D}^{(d_2)}$.} \label{Dual matter
surfaces 2-2}
\end{center}
\end{table}

We can calculate the ramification divisors for the $(2,2)$
factorization and obtain
\begin{eqnarray}
&&\widehat{r}^{(d_1)}= [\mathcal{D}^{(d_1)}]\cdot\pi^{\ast}[3c_1(B_3)-2B_2-\widehat{\xi}_2],\nonumber\\
&&\widehat{r}^{(d_2)}=
[\mathcal{D}^{(d_2)}]\cdot\pi^{\ast}[c_1(B_3)-B_2+\widehat{\xi}_2],
\end{eqnarray}
where $\widehat{r}^{(d_1)}$ and $\widehat{r}^{(d_2)}$ are the
ramification divisors for the cover $\mathcal{D}^{(d_1)}$ and
$\mathcal{D}^{(d_2)}$, respectively. We then define traceless
cover fluxes $\widehat{\gamma}^{(d_1)}_0$ and
$\widehat{\gamma}^{(d_2)}_0$ by
\begin{eqnarray}
&&\widehat{\gamma}^{(d_1)}_0=(2-p_{\mathcal{D}^{(d_1)}}^{\ast}
p_{\mathcal{D}^{(d_1)}\ast})\widehat{\gamma}^{(d_1)}
=[\mathcal{D}^{(d_1)}]\cdot \{ 2
\sigma-\pi^{\ast}[2c_1(B_3)-3B_2-\widehat{\xi}_2]\}, \nonumber\\
&&\widehat{\gamma}^{(d_2)}_0=(2-p_{\mathcal{D}^{(d_2)}}^{\ast}
p_{\mathcal{D}^{(d_2)}\ast})\widehat{\gamma}^{(d_2)}
=[\mathcal{D}^{(d_2)}]\cdot
[2\sigma+\pi^{\ast}(2B_2-\widehat{\xi}_2)],
\end{eqnarray}
where $\widehat{\gamma}^{(d_1)}$ and $\widehat{\gamma}^{(d_2)}$
are non-traceless fluxes and defined by
\begin{equation}
\widehat{\gamma}^{(d_1)}=[\mathcal{D}^{(d_1)}]\cdot \sigma,~~
\widehat{\gamma}^{(d_2)}=[\mathcal{D}^{(d_2)}]\cdot \sigma.
\end{equation}
We summarize the restriction of the fluxes to each factorized
curve in Table \ref{Chirality global 2-2 gamma}.
\begin{table}[h]
\begin{center}
\renewcommand{\arraystretch}{.75}
\begin{tabular}{|c|c|c|c|} \hline
 & $\widehat{\gamma}^{(d_2)}_0$ & $\widehat{\gamma}^{(d_1)}_0$
\\\hline

${\bf 16}^{(d_2)}$ & $-\xi_2\cdot_{B_2}(2c_1+\xi_2)$ & 0 \\ \hline

${\bf 16}^{(d_1)}$ & 0 &
$-(4c_1-t-\xi_2)\cdot_{B_2}(2c_1-t-\xi_2)$
\\\hline

${\bf 10}^{(d_2)(d_2)}$ & 0 & 0 \\ \hline

${\bf 10}^{(d_1)(d_2)}$ & 0 &
$-2(2c_1-t-2\xi_2)\cdot_{B_2}(2c_1-t-\xi_2)$
\\\hline

${\bf 10}^{(d_1)(d_1)}$ & 0 &
$2(2c_1-t-2\xi_2)\cdot_{B_2}(2c_1-t-\xi_2)$
\\\hline
\end{tabular}
\caption{Chirality induced by the fluxes
$\widehat{\gamma}^{(d_1)}_0$ and $\widehat{\gamma}^{(d_2)}_0$.}
\label{Chirality global 2-2 gamma}
\end{center}
\end{table}
We also can define two fluxes
\begin{eqnarray}
&&\widehat{\delta}^{(d_1)}=(2-p_{\mathcal{D}^{(d_2)}}^{\ast}
p_{\mathcal{D}^{(d_1)}\ast}) \widehat{\gamma}^{(d_1)}
=[\mathcal{D}^{(d_1)}]\cdot
2\sigma- [\mathcal{C}^{(d_2)}]\cdot\pi^{\ast}[2c_1(B_3)-B_2-\widehat{\xi}_2], \nonumber \\
&&\widehat{\delta}^{(d_2)}=(2-p_{\mathcal{D}^{(d_1)}}^{\ast}
p_{\mathcal{D}^{(d_2)}\ast}) \widehat{\gamma}^{(d_2)}
=[\mathcal{D}^{(d_2)}]\cdot 2\sigma-[\mathcal{C}^{(d_1)}]\cdot
\pi^{\ast}\widehat{\xi}_2.
\end{eqnarray}
Another flux we can include is \cite{Caltech:global03}
\begin{equation}
\widehat{{\rho}}^{(2,2)}=(p_{\mathcal{D}^{(d_2)}}^{\ast}-p_{\mathcal{D}^{(d_1)}}^{\ast})\widehat{\rho},
\end{equation}
where $\widehat{\rho}\in H_2(B_3,\mathbb{R})$ with
$\widehat{\rho}|_{B_2}=\rho$. We summarize the restriction of the
fluxes $\widehat{\delta}^{(d_1)}$, $\widehat{\delta}^{(d_2)}$, and
$\widehat{{\rho}}$ to each factorized curve in Table
\ref{Chirality global 2-2 delta rho}.
\begin{table}[h]
\begin{center}
\renewcommand{\arraystretch}{.75}
\begin{tabular}{|c|c|c|c|} \hline
& $\widehat{\delta}^{(d_2)}$ & $\widehat{\delta}^{(d_1)}$ &
$\widehat{{\rho}}^{(2,2)}$
\\\hline

${{\bf 16}^{(d_2)}}$ & $-2c_1\cdot_{B_2}\xi_2$ &
$-\xi_2\cdot_{B_2} (2c_1-t-\xi_2)$ & $\rho\cdot_{B_2} \xi_2$ \\
\hline

${{\bf 16}^{(d_1)}}$ & $-\xi_2\cdot_{B_2} (2c_1-t-\xi_2)$ &
$-2c_1\cdot_{B_2}(2c_1-t-\xi_2)$ & $-\rho\cdot_{B_2}
(2c_1-t-\xi_2)$
\\\hline

${{\bf 10}^{(d_2)(d_2)}}$ & $2\xi_2\cdot_{B_2} (c_1+\xi_2)$ &
$-2(c_1+\xi_2)\cdot_{B_2}(2c_1-t-\xi_2)$ & $2\rho\cdot_{B_2}(c_1+\xi_2)$\\
\hline

${{\bf 10}^{(d_1)(d_2)}}$ & 0 &
$-2(2c_1-t-2\xi_2)\cdot_{B_2}(2c_1-t-\xi_2)$ & 0
\\\hline

${{\bf 10}^{(d_1)(d_1)}}$ & $-2\xi_2\cdot_{B_2}(c_1+\xi_2)$ &
$2(3c_1-t-\xi_2)\cdot_{B_2}(2c_1-t-\xi_2)$ & $-2\rho\cdot_{B_2}
(c_1+\xi_2)$
\\\hline
\end{tabular}
\caption{Chirality induced by the fluxes
$\widehat{\delta}^{(d_1)}$, $\widehat{\delta}^{(d_2)}$, and
$\widehat{{\rho}}^{(2,2)}$.} \label{Chirality global 2-2 delta
rho}
\end{center}
\end{table}

Again we set the universal flux to be
\begin{equation}
\widehat{\Gamma}=\widetilde{k}_{d_1}
\widehat{\gamma}^{(d_1)}_0+\widehat{k}_{d_2}\widehat{\gamma}^{(d_2)}_0
+\widehat{m}_{d_1} \widehat{\delta}^{(d_1)} +\widehat{m}_{d_2}
\widehat{\delta}^{(d_2)} +
\widehat{{\rho}}=\widehat{\Gamma}^{(d_1)}+\widehat{\Gamma}^{(d_2)},
\end{equation}
where
\begin{equation}
\widehat{\Gamma}^{(d_1)}=[\mathcal{D}^{(d_1)}]\cdot
\left\{2(\widetilde{k}_{d_1}+\widetilde{m}_{d_1})\sigma
-\pi^{\ast}[2\widetilde{k}_{d_1}c_1(B_3)-(3\widetilde{k}_{d_1}+2\widetilde{m}_{d_1})B_2+(\widetilde{m}_{d_2}-\widetilde{k}_{d_1})\widehat{\xi}_2+\widehat{\rho}]\right\},
\end{equation}
\begin{equation}
\widehat{\Gamma}^{(d_2)}=[\mathcal{D}^{(d_2)}]\cdot
\left\{2(\widetilde{k}_{d_2}+\widetilde{m}_{d_2})\sigma
-\pi^{\ast}[2\widetilde{m}_{d_1}c_1(B_3)-(2\widetilde{k}_{d_2}+3\widetilde{m}_{d_2})B_2+(\widetilde{k}_{d_2}-\widetilde{m}_{d_1})\widehat{\xi}_2-\widehat{\rho}]
\right\} .
\end{equation}
Note that
\begin{eqnarray}
&&p_{\mathcal{D}^{(d_1)}\ast} \widehat{\Gamma}^{(d_1)}
=4\widetilde{m}_1{d_1}c_1(B_3)-2\widetilde{m}_{d_1}B_2-2(\widetilde{m}_{d_2}+\widetilde{m}_{d_1})\widehat{\xi}_2-2\widehat{\rho},\\
&&p_{\mathcal{D}^{(d_2)}\ast} \widehat{\Gamma}^{(d_2)} =
-4\widetilde{m}_1{d_1}c_1(B_3)+2\widetilde{m}_{d_1}B_2+2(\widetilde{m}_{d_2}-\widetilde{m}_{d_1})\widehat{\xi}_2+2\widehat{\rho}.
\end{eqnarray}
It is easy to see that $\widehat{\Gamma}^{(d_1)}$ and
$\widehat{\Gamma}^{(d_2)}$ satisfy the traceless condition
$p_{\mathcal{D}^{(d_1)}\ast}
\widehat{\Gamma}^{(d_1)}+p_{\mathcal{D}^{(d_2)}\ast}
\widehat{\Gamma}^{(d_2)}=0$. In this case the quantization
conditions are given by
\begin{equation}
\{2(\widetilde{k}_{d_1}+\widetilde{m}_{d_1})\sigma
-\pi^{\ast}[2(\widetilde{k}_{d_1}-\frac{3}{2})c_1(B_3)-(5\widetilde{k}_{d_1}+4\widetilde{m}_{d_1}+1)B_2+(\widetilde{m}_{d_2}
-\widetilde{k}_{d_1}+\frac{1}{2})\widehat{\xi}_2+\widehat{\rho}]\}
\in H_4(Z_4,\mathbb{Z}),
\end{equation}
\begin{equation}
\{2(\widetilde{k}_{d_2}+\widetilde{m}_{d_2})\sigma
-\pi^{\ast}[2\widetilde{m}_{d_1}c_1(B_3)-(4\widetilde{k}_{d_2}+5\widetilde{m}_{d_2}+1)B_2+(\widetilde{k}_{d_2}-\widetilde{m}_{d_1}-\frac{1}{2})\widehat{\xi}_2-\widehat{\rho}]
\}\in H_4(Z_4,\mathbb{Z}).
\end{equation}

\section{Conclusions}

In this paper we construct an $SU(4)$ spectral divisor of
F-theory compactified on an elliptically fibered Calabi-Yau
fourfold by using heterotic/F-theory duality.  We also explicitly
calculate the net chirality of matter fields $\bf 16$ and $\bf 10$
by using the net chirality formula Eq.~(\ref{general chirality}).
We then found agreement between the computations in F-theory framework and
in dual heterotic string. It was argued
in\cite{Marsano:2010ix} that the net chirality formula does not
depend on heterotic/F-theory duality and would be intrinsic to
F-theory. Therefore, this formula would be applicable to the cases
of F-theory compactifications without heterotic duals and the
spectral divisors can be regarded as the global completion of
semi-local spectral covers. To verify the validity of the net
chirality formula, we construct an $SU(4)$ spectral divisor in
F-theory geometry with no heterotic dual. By using this spectral
divisor and net chirality formula Eq.~(\ref{general chirality}),
we calculate the net chirality of matter fields $\bf 16$ and $\bf
10$. It turns out that the computations agree with the analysis of
the semi-local $SU(4)$ spectral cover.

To obtain realistic models, we also consider $(3,1)$ and $(2,2)$
factorizations of the $SU(4)$ spectral divisor. The explicit computation
of chiral spectra shows that the net chirality formula can be
applied to the factorized spectral divisors. By comparing with the
spectra calculated by using semi-local spectral covers, we again
found agreement between the computation in factored spectral
divisors and in factored spectral divisors. Our computations
provide an example for the validity of the spectral divisor
construction and net chirality formula. In heterotic
compactifications, the net chirality formula can be recast as an
index on a Calabi-Yau threefold. More precisely, it can be
expressed as an integral of the third Chern class of a stable
holomorphic vector bundle on the Calabi-Yau threefold. It would be
interesting to lift the net chirality formula in F-theory
framework to an index on a Calabi-Yau fourfold. The structure of
the net chirality formula should shed light on the geometry of
F-theory compactification and the nature of heterotic/F-theory
duality.

\renewcommand{\thesection}{}
\section{\hspace{-1cm} Acknowledgments}

I gratefully acknowledge hospitality and support from the Summer
School on Mathematical String Theory at Virginia Tech and PiTP
2010 Summer Program at the Institute for Advanced Study. This work
is supported in part by the NSF under grant PHY-0555575 and by
Texas A\&M University.


\renewcommand{\theequation}{\thesection.\arabic{equation}}
\setcounter{equation}{0}


\end{document}